\documentclass[aps,prd,preprint,superscriptaddress,showpacs,showkeys]{revtex4}
\usepackage{graphicx}
\usepackage{amsfonts,amsmath,amssymb,bm}

\begin{document}

\title{Divergence-type nonlinear conformal hydrodynamics}

\author{J. Peralta-Ramos}
\email{jperalta@df.uba.ar}

\affiliation{CONICET and Departamento de F\'isica, Facultad de Ciencias Exactas y Naturales, Universidad de Buenos Aires-Ciudad Universitaria, Pabell\'on I, 1428 Buenos Aires, Argentina}

\author{E. Calzetta}
\email{calzetta@df.uba.ar}
\affiliation{CONICET and Departamento de F\'isica, Facultad de Ciencias Exactas y Naturales, Universidad de Buenos Aires-Ciudad Universitaria, Pabell\'on I, 1428 Buenos Aires, Argentina}

\date{\today}

\begin{abstract}
Within the theoretical framework of divergence-type theories (DTTs), we set up a consistent nonlinear hydrodynamical description of a conformal fluid in flat space-time. DTTs go beyond second-order (in velocity gradients) theories, and are closed in the sense that they do not rely on adiabatic expansions. We show that the stress-energy tensor constructed from second-order conformal invariants is obtained from the DTT by a consistent adiabatic expansion. The DTT satisfies the Second Law, and is causal in a set of fluid states near equilibrium. Finally, we compare, analytically and numerically, the equations of motion of the DTT and its truncation to second-order terms for the case of boost invariant flow. Our numerical results indicate that the relaxation towards ideal hydrodynamics is significantly faster in the DTT than in the second-order theory. Not relying on a gradient expansion, our findings may be useful in the study of early-time dynamics and in the evolution of shock-waves in heavy-ion collisions.

\end{abstract}

\pacs{11.25.Hf, 05.70.Ln, 47.10.A-, 47.75.+f}
\keywords{divergence-type theory, relativistic causal hydrodynamics, conformal field}

\maketitle

\tableofcontents

\newpage

\section{Introduction and motivation}

There is currently a great interest in relativistic dissipative hydrodynamics, mainly due to its application to
the description of the hot
dense QCD matter created in the Relativistic Heavy Ion
Collider (RHIC) experiments \cite{heinz09,sh,rom09,ches,rang09,sonrev,libro,luzum,rischke,riscaus,ael,raj,muronga04,houv,baier06,betz}. The application of the AdS/CFT correspondence \cite{natsuume08,kov,song,bat,sonhydro,buchel,ches,lub,yaffe} to study strongly coupled conformal plasmas unaccessible to kinetic theory, has also fueled
considerable interest in the formal aspects of this formalism. The study of the hydrodynamic regime of conformal field theories is important since QCD is approximately conformal at high temperatures \cite{panero}.

The need for dissipative corrections in modeling heavy-ion collisions is (at least) two-fold. First, quantum uncertainty prevents the existence of a perfect fluid. Since first-order relativistic hydrodynamics \cite{eckart,ll} is known to have strong drawbacks, among them lack of stable solutions
and acausal propagation of perturbations, one should really go to second-order theories (see however Refs. \cite{koidemin,koidestab}). Second, the description of heavy-ion collisions in terms of perfect hydrodynamics works well in almost central Au+Au collisions near midrapidity, but gradually breaks down in non-central 
collisions and at forward rapidity \cite{baier06,houv,muronga04,sh,rom09,heinz09}. 

The complete second-order stress-energy tensor of a strongly coupled conformal fluid in $d=4$ space-time was given independently by Baier et al \cite{sonhydro}, and by Bhattacharyya et al \cite{bat}. Recently, Loganayagam \cite{logan} developed a very useful Weyl-covariant formalism and proposed a local entropy current consistent with the second-order $T_{\mu\nu}$ derived previously and with the Second Law (see also related work of Romatschke, Ref. \cite{romentropy09}). One of the most important results of Refs. \cite{sonhydro,bat,logan} is that these works show that the hydrodynamic description of a conformal fluid does not belong to the conventional Israel-Stewart \cite{israel,israelnoto,muronga04,rischke,rom09} formalism (see also Refs. \cite{rom09,ches} for a discussion of this issue). This is because the conventional (or entropy-wise) IS theory, not being a controlled gradient expansion \cite{houv,rom09}, cannot account for shear-shear coupling, which is present in the stress-energy tensor of the conformal fluid \cite{sonhydro,rom09,romentropy09,luzum} (see Refs. \cite{baier06,betz,houv} for interesting discussions on this and related issues in the context of dissipative fluid dynamics as derived from kinetic theory).

The main purpose of this paper is to set up a consistent hydrodynamical description of a conformal field theory within the theoretical framework of divergence-type theories (DTTs) \cite{geroch,geraof,liu,marc,nagy,calz98,marcprd}. Our goal is to go beyond second order theories (in velocity gradients) by formulating a theory in closed form, that is, without reliance on adiabatic expansions. We will not tackle the full problem of a conformal fluid in curved space-time \cite{sonhydro,bat,logan,romentropy09}, but limit ourselves to Minkowski space-time.
Another aim of this work is to analyze the causality properties of the DTT developed here, and to compare our results for the stress-energy tensor with those obtained from the derivative expansions of Ref. \cite{bat,sonhydro,logan}. We note that we do not calculate transport coefficients in this paper, but assume they are known either via kinetic theory or the AdS/CFT correspondence.

DTTs are interesting alternatives to the IS formalism (although they may be physically equivalent in certain cases) because the conditions for hyperbolicity and causality of the full {\it nonlinear} evolution can be stated in very simple terms. As clearly shown by Liu, Muller and Ruggeri \cite{liu}, DTTs are often more general and flexible than the IS theory, allowing a systematic derivation of nonlinear terms in constitutive equations, which are not captured by entropy-wise IS theory (see Refs. \cite{rom09,betz,houv,baier06,sonhydro,lub,rischke}). Besides, they have the extra advantage that, being the equations of motion of divergence type, discontinuous solutions (shocks) can be given mathematical meaning, which is relevant to the phenomenon of conical flow in heavy-ion collisions \cite{lub,yaffe}. Moreover, and this is an important point for what follows, the symmetries of the theory can be coded directly on the generating function of the DTT.

The main results we arrive at are that: (i) for the case in which the second-order transport coefficients $\lambda_2$ and $\lambda_3$ vanish, the second-order (in velocity gradients) stress-energy tensor constructed from conformal invariants \cite{sonhydro,logan,bat}, can be consistently derived via an adiabatic expansion from the DTT we set up; and (ii) the DTT and its adiabatic expansion are causal for states near equilibrium, and satisfy the Second Law. We also obtain, as a simple illustration, the hydrodynamic equations of the DTT for the case of boost invariant flow, and compare them to those of the second-order theory. Our numerical results show that the DTT approaches the ideal fluid behaviour faster than the second-order theory. 
 
We believe that the DTT presented here may be useful in the study of two aspects of heavy-ion collisions, both of which seem to require theories going beyond second-order velocity gradients. First, early-time dynamics, where velocity gradients are not small and for which even IS formalism shows unphysical behaviour such as reheating \cite{houv} (see also  Chesler and Yaffe \cite{yaffe}, who study the creation and evolution of a boost invariant anisotropic plasma directly from the gravity side, and El, Xu and Greiner \cite{ael}, who develop a novel third order theory). Second, the evolution of initial state fluctuations \cite{calz98}, for which higher order terms are crucial (see particularly the work of Lublinsky and Shuryak \cite{lub}, who developed a linearized hydrodynamical theory that includes, in principle, all-order velocity gradients). In this respect (that of ``resumming" higher order velocity gradients), the developments of Ref. \cite{lub} are related to ours. We note that the DTT developed here containts, in addition, all quadratic terms in velocity.

The paper is organized as follows. In Section \ref{app1}, we briefly review divergence-type theories. In Section \ref{main} we first review some basic properties of the hydrodynamics of conformal field theories, and then set up the divergence-type theory of a conformal fluid. We also prove that the DTT satisfies the Second Law exactly, and obtain the hydrodynamic equations.
In Section \ref{caus} we investigate the causality properties of the DTT, for fluid states near equilibrium. In Section \ref{eqsmotion}, we show that the dissipative part of the stress-energy tensor obtained from second-order (in velocity gradients) conformal invariants can be obtained from a consistent adiabatic expansion of the DTT. We note that the DTT cannot reproduce terms containing the vorticity tensor. In Section \ref{boost} we compare, analytically and numerically, the hydrodynamic equations of the DTT and of the second-order theory for the case of Bjorken flow. The paper closes up with a brief summary of results.

\section{Divergence-type theories}
\label{app1}
In this section we give a brief summary of divergence-type theories (DTTs). Detailed discussions can be found in Refs. \cite{geroch,calz98,marc,marcprd} (see also Ref. \cite{liu}).

According to Geroch and Lindblom \cite{geroch}, the hydrodynamical description of a nonequilibrium state requires, besides the particle current $N_a$ and the stress-energy tensor $T_{\mu\nu}$, a new third order tensor $A_{\mu\nu\rho}$ obeying an equation of motion of divergence type. The dynamical equations are the conservation laws of $N_\mu$ and $T_{\mu\nu}$, together with an equation describing the dissipative part:
\begin{equation}
A^{\mu\nu\rho}_{~~~;\mu} = I^{\nu\rho}
\label{eomapp}
\end{equation}
where $A$ and $I$ are algebraic local functions of $N$ and $T$ and symmetric in the indices $(\nu,\rho)$. A semicolon stands for a covariant derivative. The entropy current is extended to
\begin{equation}
S^\mu = \Phi^\mu - \beta_\nu T^{\mu\nu} - \alpha N^\mu - A^{\mu\nu\rho} \xi_{\nu\rho}
\end{equation}
where $\beta_\nu=u_\nu/T$ is the temperature vector, $\alpha=\mu/T$ is the affinity, $\Phi^\mu$ is the thermodynamic potential and $\xi_{\nu\rho}$ is symmetric, traceless and vanish in equilibrium. Note that, in equilibrium, $\beta^\mu$ is Killing and $\alpha$ is constant.

We now require that the entropy and the thermodynamical potential be algebraic functions of $(\alpha,\beta_\mu,\xi_{\mu\nu})$. If the entropy production is to be nonnegative, then
\begin{equation}
\frac{\partial \Phi^\mu}{\partial \alpha} = N^\mu; \qquad
\frac{\partial \Phi^\mu}{\partial \beta_\nu} = T^{\mu\nu}; \qquad
\frac{\partial \Phi^\mu}{\partial \xi_{\nu\rho}} = A^{\mu\nu\rho}
\end{equation}
Thus, as a consequence of the equations of motion, the entropy production rate is 
\begin{equation}
S^\mu_{;\mu} = -I^{\nu\rho}\xi_{\nu\rho} ~. 
\end{equation}

Since the stress-energy tensor is symmetric, we must also have
\begin{equation}
\Phi^\mu=\frac{\partial \chi}{\partial \beta_\mu}
\end{equation}
where $\chi(\alpha,\beta_\mu,\xi_{\mu\nu})$ is the so-called generating function of the theory. This means that every DTT is completely determined once $\chi$ and $I$ are specified as algebraic functions of $\alpha,\beta_\mu,\xi_{\mu\nu}$. The theory thus constructed satisfies the principles of relativity and entropy, and fully exploits the latter \cite{liu}.

Introducing the symbol $\zeta^A$ to denote the set $(\alpha,\beta_\mu,\xi_{\mu\nu})$, $A^\mu_B$ the set $(N^\mu,T^{\mu\nu},A^{\mu\nu\rho})$ and $I_B$ the set $(0,0,I_{\mu\nu})$, the theory is summed up in the equations
\begin{equation}
\begin{split}
A^\mu_B &= \frac{\partial \Phi^\mu}{\partial \zeta^B} \\
S^\mu_{;\mu} &= -I_B \zeta^B \\
A^\mu_{B;\mu} &= I_B ~~.
\end{split}
\end{equation}
The equations of motion can also be written as
\begin{equation}
M^\mu_{BC} \zeta^C_{;\mu} = I_B
\label{MI}
\end{equation}
where
\begin{equation}
M^\mu_{BC}=M^\mu_{CB}=\partial^2\Phi^\mu/\partial \zeta^B \partial\zeta^C ~~.
\end{equation}
The system of equations (\ref{MI}) is automatically symmetric since matrix $M^\mu_{BC}$ is symmetric in the indices $A$ and $B$ due to the fact that partial derivatives conmute.
Causality is therefore ensured if the quadratic form $M^\mu_{BC}\omega_\mu$ is negative definite for all future directed timelike vectors $\omega_\mu$, or equivalently if $Q^\mu=M^\mu_{BC}\delta \zeta^B \delta \zeta^C$ is timelike and future oriented for any displacement $\delta \zeta^A$ from an equilibrium state \cite{geroch,geraof}. Note that causality depends on the form of $M^\mu_{BC}$ and not of $I_B$.

\section{Conformal hydrodynamics as a divergence-type theory}
\label{main}

We consider the hydrodynamic regime of a conformal quantum field theory in $d=4$ flat space-time (for a general discussion, not limited to conformal fields, see Ref. \cite{calz00,libro}). In such a theory, the classical action evaluated on the classical equations of motion is invariant under a Weyl transformation $g_{\mu \nu}\rightarrow e^{-2\omega(x^\gamma)}g_{\mu\nu}$, where $\omega$ is a function of space-time coordinates $x^\gamma$. The classical stress-energy tensor of such a theory is necessarily traceless, while the quantum one presents Weyl anomaly. As shown in Ref. \cite{sonhydro}, in even dimensions $d$ the number of derivatives appearing in the Weyl anomaly is precisely $d$, which means that second-order hydrodynamics in $d=4$ dimensions {\it is} Weyl invariant. It is straightforward to show that, for a conformal theory,
\begin{equation}
T^{\mu\nu}\rightarrow e^{(d+2)\omega}T^{\mu\nu}
\label{transT}
\end{equation}
under a Weyl transformation. Therefore, for a conformal fluid the conservation law of energy-momentum 
$T^{\mu\nu}_{; \nu}=0$, where a semi-colon denotes covariant differentiation, is automatically Weyl covariant (see, for example, Refs. \cite{logan,boul,rang09}).

The tracelessness condition $T^\mu_\mu=0$ imposes $\rho=(d-1)p$ and $\zeta=0$, where $\rho$ is the energy density in the local frame, $p$ is the thermodynamic pressure, and $\zeta$ is the bulk viscosity. The transformation rule for $T^{\mu\nu}$ implies $\rho \rightarrow e^{d\omega} \rho$, the four-velocity $u^\mu\rightarrow e^{\omega} u^\mu$ and the temperature $T\rightarrow e^{\omega} T$, which means that the temperature vector $\beta^\mu=u^\mu/T$ has conformal weight equal to zero. In addition, Eq. (\ref{transT}) implies that the shear viscosity $\eta = A T^{d-1}$, with $A$ a constant (see, for instance, Refs. \cite{sonhydro,logan,bat,romentropy09}). In equilibrium $\beta_{(\mu;\nu)}=0$, which means that $\beta_\mu$ is a Killing vector (parenthesis around indices stand for symmetrization). The four-velocity is normalized as $u_\mu u^\mu=-1$, and we use the signature $(-,+,+,+)$. We will make use of these properties in what follows.

A divergence-type theory is completely specified by its generating function $\chi(\alpha,\beta^\mu,\xi^{\mu\nu})$ and source tensor $I_{\rho\gamma}(\alpha,\beta^\mu,\xi^{\mu\nu})$, where $(\alpha,\beta^\mu,\xi^{\mu\nu})$ are the fugacity, the temperature vector and a symmetric and traceless tensor that vanish in equilibrium, respectively. Our starting point will therefore be the specification of $\chi$ and $I$ as algebraic functions of the set $(\alpha,\beta^\mu,\xi^{\mu\nu})$. We will deal with the source tensor later on, so for the moment we focus on $\chi$.

Since we want to construct a quadratic DTT (in deviations from equilibrium), we will consider terms which are at most quadratic in the nonequilibrium tensor $\xi^{\mu\nu}$. For simplicity, we will restrict ourselves to a conformal theory with no conserved charges, which means $\alpha=0$. In this case, it is convenient to employ the energy or Landau-Lifshitz frame \cite{libro,ll,muronga04}. The generating function which satisfies these requirements can be written as
\begin{equation}
\begin{split}
\chi &= \chi_{(0)} + \chi_{(1)} + \chi_{(2)} \\
&= \chi_0(T) + \chi_1(T)\xi_{\mu\nu}u^\mu u^\nu + \sum_{i=1}^3 \chi_2^{(i)}(T) S_{(i)}^{\mu\nu\rho\sigma}\xi_{\mu\nu}\xi_{\rho\sigma}
\end{split}
\label{genchi}
\end{equation}
with
\begin{equation}
\begin{split}
S_{(1)}^{\mu\nu\rho\sigma} &= \Delta^{\mu(\rho}\Delta^{\sigma)\nu}-\frac{1}{3}\Delta^{\mu\nu}\Delta^{\rho\sigma} \\
S_{(2)}^{\mu\nu\rho\sigma} &= u^{(\mu}\Delta^{\nu)(\rho}u^{\sigma)} \\
S_{(3)}^{\mu\nu\rho\sigma} &= \frac{3}{4}\bigg(\frac{\Delta^{\mu\nu}}{3}+u^\mu u^\nu \bigg) \frac{3}{4}\bigg(\frac{\Delta^{\rho\sigma}}{3}+u^\rho u^\sigma \bigg) \\
\Delta^{\mu\nu} &= g^{\mu\nu} +u^\mu u^\nu ~~ .
\end{split}
\label{proj}
\end{equation}
This is the most general local scalar constructed from $T$, $g_{\mu\nu}$, $u^\mu$ and $\xi_{\mu\nu}$, which is quadratic in the latter. In Eq. (\ref{proj}), $\Delta^{\mu\nu}$ is the spatial projector orthogonal to $u^\mu$. From the transformation rule for $g^{\mu\nu}$ and $u^\mu$ we immediately obtain $\Delta^{\mu\nu}\rightarrow e^{2\omega}\Delta^{\mu\nu}$. The tensors $S_{(i)}^{\mu\nu\rho\sigma}$ produce the most general decomposition of a symmetric and traceless tensor around a timelike direction $u^\mu$.

As already mentioned, conformal invariance requires $T^{\mu\nu}\rightarrow e^{(d+2)\omega}T^{\mu\nu}$ and $T^\mu_\mu=0$. In a DTT, this means
\begin{equation}
\frac{\partial^2 \chi}{\partial \beta_\mu \partial \beta_\nu} \rightarrow e^{(d+2)\omega} \frac{\partial^2 \chi}{\partial \beta_\mu \partial \beta_\nu}
\label{ci-dtt1}
\end{equation}
and
\begin{equation}
g_{\mu\nu}\frac{\partial^2 \chi}{\partial \beta_\mu \partial \beta_\nu} = 0  ~~.
\label{ci-dtt2}
\end{equation}
Note that 
\begin{equation}
\begin{split}
\frac{\partial u^\mu}{\partial \beta_\nu} &= T \Delta^{\mu\nu} \\
\frac{\partial \Delta^{\alpha\gamma}}{\partial \beta_\nu} &= T^2 [\Delta^{\alpha\nu}\beta^\gamma + \beta^\alpha\Delta^{\gamma\nu}] ~~ \textrm{and} \\
\frac{\partial T}{\partial \beta_\nu} &= T^2 u^{\nu} ~~.
\end{split}
\label{derivbeta} 
\end{equation}

In the following, we will describe how these conditions determine the scalar functions $\chi_{0,1}$ and $\chi_2^{(i)}$ appearing in Eq. (\ref{genchi}). It is clear that the conditions given in Eqs. (\ref{ci-dtt1}) and (\ref{ci-dtt2}) must be satisfied separately by the zeroth, first and second order terms of the expansion of $\chi$. This is because truncating the expansion at zeroth, first and second order does not break the conformal invariance of the resulting hydrodynamic theory. In other words, the zeroth, first and second order terms of $T^{\mu\nu}$ are independent of each other. 

\subsection{Perfect fluid}

At zeroth-order, condition (\ref{ci-dtt2}) on the stress-energy tensor
\begin{equation}
T^{\mu\nu}_0=pg^{\mu\nu}+u^\mu u^\nu[p+\rho]
\label{tperf}
\end{equation}
implies $\rho=(d-1)p$. In a DTT we have (recall that $\rho$ and $p$ are {\it equilibrium} quantities and thus completely determined by $\chi_0$)
\begin{equation}
T^{\mu\nu}_0=T^3 g^{\mu\nu}\frac{d \chi_0}{d T} + T^3 u^\mu u^\nu \bigg(3\frac{d \chi_0}{d T}+T\frac{d^2 \chi_0}{d T^2}\bigg) ~,
\end{equation}
which implies
\begin{equation}
\begin{split}
p &= T^3\frac{d \chi_0}{d T} ~~\textrm{and} \\
\rho &= T^3\bigg(2 \frac{d \chi_0}{d T} + T\frac{d^2 \chi_0}{d T^2}\bigg) ~~ .
\end{split}
\label{pandrho}
\end{equation}
Therefore, we find that in a conformal DTT $\chi_0$ must satisfy
\begin{equation}
\frac{d^2 \chi_0}{d T^2} = \frac{(d-3)}{T}\frac{d \chi_0}{d T} ~~.
\label{diffchi0}
\end{equation}
The solution to Eq. (\ref{diffchi0}) is
\begin{equation}
\chi_0 = aT^{d-2} + a'
\label{chi0final}
\end{equation}
where $a$ and $a'$ are constants. Note that $a'$ is irrelevant since it does not change $T_0^{\mu\nu}$, so we set it to zero.

From the transformation rule for $T$ we immediately obtain $\chi_0 \rightarrow e^{(d-2)\omega}\chi_0$.
Taking into account that $\beta^\alpha$ is Weyl invariant while
\begin{equation}
\beta_\alpha \rightarrow \frac{e^{-\omega}u_\alpha}{e^\omega T} = e^{-2\omega} \beta_\alpha
\end{equation}
we have
\begin{equation}
\begin{split}
T_0^{\mu\nu}=\frac{\partial^2 \chi_0}{\partial\beta_\mu \partial \beta_\nu} &\rightarrow e^{(d-2)\omega}e^{4\omega} T_0^{\mu\nu} \\
&= e^{(d+2)\omega} T_0^{\mu\nu}
\end{split}
\end{equation}
as it should.

From the above it is clear that, in order to obtain the correct conformal weight for the complete second-order stress-energy tensor (Eq. (\ref{ci-dtt1})), the generating function given in Eq. (\ref{genchi}) must transform like $\chi_0$ under a Weyl transformation. That is, we require that
\begin{equation}
\chi \rightarrow e^{(d-2)\omega} \chi ~~ .
\label{totchi}
\end{equation}
As already noted, this implies that the first and second order terms $\chi_{(1)}$ and $\chi_{(2)}$ have conformal weight equal to $(d-2)$.

\subsection{Linear DTT}
\label{firstsec}

We will now determine $\chi_1(\lambda)$ and from it the stress-energy tensor at first order in the nonequilibrium tensor $\xi_{\alpha\beta}$:
\begin{equation}
T^{\mu\nu}=T_0^{\mu\nu}+\tau_1^{\mu\nu}
\end{equation}
with
\begin{equation}
\tau^{\mu\nu}_1 = \frac{\partial^2 \chi_{(1)}}{\partial \beta_\mu \partial \beta_\nu} ~~ .
\end{equation}
It is convenient to rewrite the first order term in $\chi$ as
\begin{equation}
\chi_{(1)}=\chi_1 \xi_{\rho\sigma} u^\rho u^\sigma = \tilde{\chi}_1 \xi_{\rho\sigma} \beta^\rho \beta^\sigma
\label{chi1bar}
\end{equation}
with $\tilde{\chi}_1=T^2 \chi_1$. 
The tracelessness condition $g_{\mu\nu}\tau^{\mu\nu}_1=0$ implies the following differential equation for $\tilde{\chi}_1$:
\begin{equation}
T \frac{d^2 \tilde{\chi}_1}{dT^2} - 2\frac{d \tilde{\chi}_1}{dT} = 0
\label{diffeq1}
\end{equation}
whose solution is
\begin{equation}
\tilde{\chi}_1 (T)= b + c T^3
\label{chi1final}
\end{equation}
with $b$ and $c$ constant.  

The issue of how to choose the integration constants $b$ and $c$ is not trivial. 
The criterium of a bounded solution is not compelling enough\footnote{We thank the referee for pointing out this fact to us.} to
support the choice of constant $c=0$. However, we would like to point out
that the requirement that the generating function must have a definite
transformation law under conformal transformations means that the two
constants $b$ and $c$ cannot be both different from zero at the same time.
Therefore there are two families of conformal divergence type theories: one with $b=0$, $c\neq 0$ and the other with $b \neq 0$ and $c=0$. Moreover,
the physical content of both familes is the same. Indeed, if we substitute
$b$ by $cT^3$, but after computing the energy-momentum tensor we replace
$\xi_{\mu\nu}$ by $T^{-3}\xi_{\mu\nu}$ we obtain once again Eq. (\ref{xiuno}). Of
course, to reach this conclusion we use that the physical nonequilibrium
tensor $\xi_{\mu\nu}$ is transverse with respect to the four-temperature
$\beta^{\mu}$ (this fact will be shown in what follows), so extra terms in the energy-momentum tensor vanish
identically. 
Given the physical equivalence of both classes of theories, we have chosen
to investigate only the $b\neq 0$ case as a matter of simplicity.

As it can be seen from Eqs. (\ref{totchi}) and (\ref{chi1bar}) and the fact that $\beta^\mu$ is Weyl invariant, setting $c=0$ implies that the conformal weight of $\xi_{\mu\nu}$ is equal to $d-2$. Therefore,
\begin{equation}
\begin{split}
\xi_{\mu\nu} &\rightarrow e^{(d-2)} \xi_{\mu\nu} ~~ \textrm{and} \\
\xi^{\mu\nu} &\rightarrow e^{(d+2)} \xi^{\mu\nu} ~~.
\end{split}
\label{transxi}
\end{equation}

From the generating functional determined above, the tensor of fluxes
\begin{equation}
A^{\delta \alpha \gamma}= \frac{\partial^2 \chi_{(1)}}{\partial \xi_{\alpha \gamma}\partial \beta_\delta}
\label{a11}
\end{equation}
becomes
\begin{equation}
A^{\delta \alpha \gamma}=b(g^{\delta\alpha}\beta^\gamma + g^{\gamma\delta}\beta^\alpha) - \frac{b}{2}\beta^\delta \delta^{\alpha\gamma} ~~,
\label{a11res}
\end{equation}
where we have used that
\begin{equation}
\frac{\delta\xi_{\rho\sigma}}{\delta \xi_{\alpha\gamma}} = \frac{1}{2}(\delta_\rho^\alpha \delta^\gamma_\sigma+\delta^\alpha_\sigma \delta_\rho^\gamma)-\frac{1}{4}g_{\rho\sigma}\delta^{\alpha\gamma}
\label{deltaxi}
\end{equation}
which follows since $\xi_{\mu\nu}$ is symmetric and traceless.

The divergence of the tensor of fluxes is
\begin{equation}
A^{\delta\alpha\gamma}_{~~~;\delta}=b\beta^{(\alpha;\gamma)} -\frac{b}{2}\beta^\delta_{;\delta}\delta^{\alpha\gamma} ~~.
\label{diva}
\end{equation}
The first order divergence-type theory is summed up by Eq. (\ref{diva}) together with
\begin{equation}
\tau_1^{\mu\nu} =  b\xi^{\mu\nu}
\label{divtau}
\end{equation}
and
\begin{equation}
A^{\delta\alpha\gamma}_{~~~;\delta} =  I^{\alpha\gamma} ~~.
\label{divi}
\end{equation}

The system of equations (\ref{diva}-\ref{divi}) must lead to the same $\tau_1^{\mu\nu}$ of Eckart's theory (for a conformal fluid), which can be written as \cite{calz98,geroch,liu}:
\begin{equation}
\tau_{1E}^{\mu\nu}= -\eta S_{(1)}^{\mu\nu\rho\sigma}\bigg(u_{(\rho;\sigma)}-\frac{T_{;\sigma)}}{T}u_{(\rho}\bigg)= -\eta T S_{(1)}^{\mu\nu\rho\sigma}\beta_{(\rho;\sigma)}
\label{taue}
\end{equation}
where the last equality follows from the transversality of $S_{(1)}^{\mu\nu\rho\sigma}$.
Note that, since
\begin{equation}
\begin{split}
\beta_{(\rho;\sigma)} &\rightarrow e^{-2\omega} \beta_{(\rho;\sigma)} \\
S_{(1)}^{\mu\nu\rho\sigma} &\rightarrow e^{4\omega}S_{(1)}^{\mu\nu\rho\sigma} ~,
\end{split}
\label{sandbetatrans}
\end{equation}
we must have, from Eq. (\ref{taue}), that $\eta \rightarrow e^{3\omega}\eta$, and therefore $\eta(T) = \textrm{const.} T^3$.

In order for the first order stress-energy tensor obtained from the DTT to coincide with that of Eckart's theory we must provide a linear relationship between the source tensor $I^{\alpha\gamma}$ and the nonequilibrium tensor $\xi^{\alpha\gamma}$:
\begin{equation}
I^{\alpha\gamma}=-D^{\alpha\gamma\rho\sigma}\xi_{\rho\sigma} ~~.
\label{ID}
\end{equation}
Using Eqs. (\ref{diva}-\ref{divi}) and Eq. (\ref{taue}) we obtain
\begin{equation}
D^{\alpha\gamma\rho\sigma} = \frac{b^2}{\eta T} S_{(1)}^{\alpha\gamma\rho\sigma} ~~,
\label{D}
\end{equation}
where we have used that
\begin{equation}
g_{\mu\nu}S_{(1)}^{\mu\nu\rho\sigma} = \Delta^{\mu(\rho}\Delta^{\sigma)}_\mu - \frac{1}{3}\Delta_\mu^\mu \Delta^{\rho\sigma} = 0
\end{equation}
since
\begin{equation}
\Delta^{\alpha\beta}\Delta^\gamma_\alpha = \Delta^{\beta \gamma} ~,~ \Delta^{\alpha\beta}=\Delta^{\beta\alpha} ~\textrm{and}~  \Delta^\mu_\mu = 3 ~.
\end{equation}

Since the conformal weights of $\xi_{\rho\sigma}$ and $I^{\alpha\gamma}$ are both equal to 2 (see Eqs. (\ref{diva})-(\ref{divi})), it is seen from Eq. (\ref{ID}) that the tensor $D^{\alpha\gamma\rho\sigma}$ must be Weyl invariant. From Eq. (\ref{D}) this implies $\eta \propto T^3$, as before. Note also that the requirement $\tau_{1E}^{\mu\nu}=\tau_1^{\mu\nu}$ automatically implies $\beta_\mu\xi^{\mu\nu}=0$, since $S_{(1)}^{\mu\nu\rho\sigma}$ is transverse. We have
\begin{equation}
\xi_{\alpha\gamma}= -\frac{\eta}{b}\sigma_{\alpha\gamma}
\label{xiuno}
\end{equation}
where $\sigma_{\alpha\gamma}=S_{(1)\alpha\gamma\mu\nu} u^{(\mu;\nu)}$. The physical meaning of $\xi^{\mu\nu}$ being transverse is that the bulk viscosity
and the heat flux, which are both proportional to $\beta^\alpha\xi_{\alpha\gamma}$ \cite{calz98}, vanish. The vanishing of the heat flux is expected since the chemical potential is zero, whereas the bulk viscosity is zero since the theory is conformal. In the next section we will show that the transversality of $\xi^{\mu\nu}$ holds in the quadratic theory as well.

\subsection{Quadratic DTT}
\label{secondsec}

We now go over to the quadratic stress-energy tensor given by
\begin{equation}
T^{\mu\nu}=T_0^{\mu\nu}+\tau_1^{\mu\nu}+\tau_2^{\mu\nu}
\end{equation}
with
\begin{equation}
\tau_2^{\mu\nu} = \frac{\partial^2 \chi_{(2)}}{\partial \beta_\mu \partial \beta_\nu} ~~.
\end{equation}

From the conformal weights of $\xi_{\mu\nu}$, given in Eq. (\ref{transxi}), and of $S_{(i)}^{\alpha\gamma\rho\sigma}$ (see Eq. (\ref{sandbetatrans})) it is seen that, to obtain the correct conformal weight for $T^{\mu\nu}$, we must have $\chi_2^{(i)}=c_i T^{-6}$, where $c_i$ are constants to be determined. This ensures that $\chi_{(2)}$ in Eq. (\ref{genchi}) has conformal weight equal to $(d-2)$, which means $\tau_2^{\mu\nu}$ has conformal weight equal to $(d+2)$.

The tracelessness condition, Eq. (\ref{ci-dtt2}), will determine relations among the coefficients $c_i$. 
The quadratic contribution to the stress-energy tensor can be written as
\begin{equation}
\tau_2^{\mu\nu} = \Gamma^{\mu\nu} \sum_{i=1}^3 c_i S_{(i)}^{\alpha\gamma\rho\sigma} \xi_{\alpha\gamma}\xi_{\rho\sigma}
\label{tau2fromGamma}
\end{equation}
where we have defined the operator
\begin{equation}
\begin{split}
\Gamma^{\mu\nu} &= 6T^{-4} (4T^2\beta^\mu \beta^\nu - \delta^{\mu\nu} ) \\
& ~ -6 T^{-4}\bigg(\beta^\nu \frac{\partial }{\partial \beta_\mu}+ \beta^\mu \frac{\partial }{\partial \beta_\nu}\bigg)\\
& ~ + T^{-6} \frac{\partial^2 }{\partial \beta_\mu \partial \beta_\nu} ~~.
\end{split}
\end{equation}
In this notation, the trace of $\tau_2^{\mu\nu}$ becomes
\begin{equation}
g_{\mu\nu}\tau_2^{\mu\nu} = \Gamma^\mu_\mu \sum_{i=1}^3 c_i S_{(i)}^{\alpha\gamma\rho\sigma} \xi_{\alpha\gamma}\xi_{\rho\sigma}
\label{tracetau2}
\end{equation}
with
\begin{equation}
\Gamma^\mu_\mu = - T^{-4}\bigg(48 + 12\beta_\mu \frac{\partial}{\partial \beta_\mu} -T^{-2}\frac{\partial^2}{\partial \beta_\mu\partial \beta^\mu}\bigg) ~.
\label{gammamumu}
\end{equation}

Computing the derivatives and equating the coefficients of the Lorentz invariants
\begin{equation}
\Delta^{\alpha\rho}\Delta^{\gamma\sigma}\xi_{\alpha\gamma}\xi_{\rho\sigma}, ~ \beta^\alpha\beta^\sigma\Delta^{\rho\gamma}\xi_{\alpha\gamma}\xi_{\rho\sigma} ~ \textrm{and} ~ \beta^\alpha\beta^\gamma\beta^\rho\beta^\sigma\xi_{\alpha\gamma}\xi_{\rho\sigma}
\end{equation}
 to zero ($g_{\mu\nu}\tau_{2}^{\mu\nu} = 0$), we find a linear system of equations for the three unknowns $c_i$. It turns out that the equations involving $c_2$ and $c_3$ (which come from the invariants $\beta^2\Delta\xi^2$ and $\beta^4\xi^2$) are inconsistent with each other, the only way out being imposing that
\begin{equation}
\beta^\alpha\xi_{\alpha\gamma}=0 ~.
\label{xiistransverse}
\end{equation}
That is, the tracelessness of $\tau_2^{\mu\nu}$ forces the transversality of the nonequilibrium tensor $\xi_{\alpha\beta}$. This means that the heat flux and the bulk viscosity remain zero at second order, which is a satisfying result. 
With this additional requirement on the nonequilibrium tensor, $c_2$ is left unspecified, while the remaining equation (coming from the invariant $\Delta^{\alpha\rho}\Delta^{\gamma\sigma}\xi_{\alpha\gamma}\xi_{\rho\sigma}$ and relating $c_1$ and $c_3$) reads
\begin{equation}
c_1 = -\frac{3}{8}c_3 ~~.
\end{equation}

Therefore, we have found that the quadratic part of generating function of a conformal fluid can be written as
\begin{equation}
\begin{split}
\chi_{(2)} &= T^{-6}\bigg[c_1 \bigg(S_{(1)}^{\alpha\gamma\rho\sigma}-\frac{8}{3}S_{(3)}^{\alpha\gamma\rho\sigma}\bigg)\\
&~ + c_2 S_{(2)}^{\alpha\gamma\rho\sigma} \bigg] \xi_{\alpha\gamma}\xi_{\rho\sigma} ~.
\end{split}
\label{chi2final}
\end{equation}

From Eq. (\ref{chi2final}) we can calculate the tensor of fluxes
\begin{equation}
A^{\delta\alpha\gamma}= A_E^{\delta\alpha\gamma} + \frac{\partial^2 \chi_{(2)}}{\partial \beta_\delta \partial \xi_{\alpha\gamma}} = A_E^{\delta\alpha\gamma} + A_2^{\delta\alpha\gamma}
\end{equation}
where $A_E^{\delta\alpha\gamma}$ is the first order term given by Eqs. (\ref{a11}) and (\ref{a11res}). We can rewrite $A_2$ as
\begin{equation}
A_2^{\delta\alpha\gamma} = G^{\delta\alpha\gamma\rho\sigma} \xi_{\rho\sigma} ~,
\end{equation}
with
\begin{equation}
\begin{split}
G^{\delta\alpha\gamma\rho\sigma} &= 2T^{-4}\bigg[ c_1\bigg(-6 \beta^\delta S_{(1)}^{\alpha\gamma\rho\sigma} +\beta^\alpha S_{(1)}^{\delta\gamma\rho\sigma} + \beta^\gamma S_{(1)}^{\alpha\delta\rho\sigma}\bigg) \\
&~ + 2c_2\Delta^{\sigma\delta}\bigg(\Delta^{\rho\alpha}\beta^\gamma+\beta^\alpha\Delta^{\rho\gamma}\bigg) \\
&~ + 2c_2 \Delta^{\rho\delta}\bigg(\Delta^{\sigma\alpha}\beta^\gamma+\beta^\alpha\Delta^{\sigma\gamma}\bigg) \bigg] ~.
\end{split}
\label{a2G}
\end{equation}

The divergence of $A^{\delta\alpha\gamma}_{2}$ is
\begin{equation}
A^{\delta\alpha\gamma}_{2 ;\delta}=\frac{\partial G^{\delta\alpha\gamma\rho\sigma}}{\partial \beta_\pi}\beta_{\pi;\delta}\xi_{\rho\sigma}+G^{\delta\alpha\gamma\rho\sigma}\frac{\partial \xi_{\rho\sigma}}{\partial \xi_{\pi\theta}}\xi_{\pi\theta;\delta} ~.
\label{symbdiva}
\end{equation}
We get, after some algebra,
\begin{equation}
\begin{split}
A^{\delta\alpha\gamma}_{2 ;\delta} &= -4T^{2} \beta^\pi G^{\delta\alpha\gamma\rho\sigma}\xi_{\rho\sigma}\beta_{\pi;\delta} \\
&~ + 2T^{-4}c_1\bigg(-6K^{\pi\delta\alpha\gamma\rho\sigma}_1+K^{\pi\alpha\delta\gamma\rho\sigma}_1 \\
&~ + K^{\pi\gamma\alpha\delta\rho\sigma}_1 \bigg) \xi_{\rho\sigma}\beta_{\pi;\delta}
+ G^{\delta\alpha\gamma\rho\sigma}\xi_{\rho\sigma;\delta} \\
&~ +4T^{-4}c_2 \bigg(\Delta^{\sigma\delta} (\Delta^{\rho\alpha}\delta^{\pi\gamma}+ \Delta^{\rho\gamma}\delta^{\pi\alpha}) \\
&~ + \Delta^{\rho\delta} (\Delta^{\sigma\alpha}\delta^{\pi\gamma}+ \Delta^{\sigma\gamma}\delta^{\pi\alpha}) \bigg)
\xi_{\rho\sigma}\beta_{\pi;\delta}
\end{split}
\label{diva2}
\end{equation}
with
\begin{equation}
K^{\pi\delta\alpha\gamma\rho\sigma}_1 = \delta^{\pi\delta}S_{(1)}^{\alpha\gamma\rho\sigma} + T^2\beta^\delta\bigg(\beta^\alpha S_{(1)}^{\pi\gamma\rho\sigma} + \beta^\gamma S_{(1)}^{\alpha\pi\rho\sigma}\bigg) ~.
\end{equation}

In order to have a complete theory at second order in deviations from equilibrium, we must find a suitable source tensor $I_2$ quadratic in $\xi^{\mu\nu}$. We will find the constraints imposed on $I_2$ by requiring that the Second Law holds, and by the fact that $\xi^{\mu\nu}$ is traceless and transverse, and find an explicit expression for $I_2$. Guided by linear results, we will consider that $I_2$ has the form
\begin{equation}
I_2^{\alpha\gamma} = J^{\alpha\gamma\rho\sigma\mu\nu}\xi_{\rho\sigma}\xi_{\mu\nu} ~,
\label{IJxi}
\end{equation}
where $J=J(\beta^{\delta},\Delta^{\delta\pi})$.

In a DTT, the entropy production is simply
$S^\mu_{; \mu}=-I^{\alpha\gamma}\xi_{\alpha\gamma}$. We have
\begin{equation}
\begin{split}
S^\mu_{; \mu} &= S^\mu_{; \mu}\bigg|_1 + S^\mu_{; \mu}\bigg|_2 = D^{\alpha\gamma\rho\sigma}\xi_{\rho\sigma}\xi_{\alpha\gamma}-I_2^{\alpha\gamma}\xi_{\alpha\gamma} \\
&~= D^{\alpha\gamma\rho\sigma}\xi_{\rho\sigma}\xi_{\alpha\gamma}-J^{\alpha\gamma\rho\sigma\mu\nu} \xi_{\alpha\gamma}\xi_{\rho\sigma}\xi_{\mu\nu} ~.
\end{split}
\end{equation}
$S^\mu_{; \mu}|_1$ is the entropy production of the linear DTT (in $\xi$), and is clearly positive definite. The problem comes from $S^\mu_{; \mu}|_2$, which has no definite sign. In order for the Second Law to hold for arbitrary $\xi$, we must require that
\begin{equation}
I_2^{\alpha\gamma} \xi_{\alpha\gamma} = 0 ~.
\label{I2xi}
\end{equation}

We will now find the explicit form of $J^{\alpha\gamma\rho\sigma\mu\nu}$ (see Eq. (\ref{IJxi})). Since $I_2^{\alpha\gamma}$ is a local function of $\beta^\mu$ and $\Delta^{\mu\nu}$, $\xi^{\mu\nu}$ is traceless and transverse, and Eq. (\ref{I2xi}) must hold, we see that $J^{\alpha\gamma\rho\sigma\mu\nu}$ can only have two terms, one proportional to $\beta^\alpha\beta^\gamma$ and the other to $\Delta^{\alpha\gamma}$. In addition, the only non-vanishing scalar we can form out of $\xi_{\rho\sigma}\xi_{\mu\nu}$ is
\begin{equation}
S_{(1)}^{\rho\sigma\mu\nu}\xi_{\rho\sigma}\xi_{\mu\nu} = \xi^{\rho\sigma}\xi_{\rho\sigma} ~.
\end{equation}
Therefore, we have
\begin{equation}
J^{\alpha\gamma\rho\sigma\mu\nu}=\bigg(f_1(T)\beta^\alpha\beta^\gamma + f_2(T)\Delta^{\alpha\gamma} \bigg)S_{(1)}^{\rho\sigma\mu\nu}
\label{jexplicit}
\end{equation}
where the $f_i(T)$ are functions of temperature we must determine. It can be checked from Eq. (\ref{diva2}) that
\begin{equation}
\beta_\alpha \beta_\gamma A_{2;\delta}^{\delta \alpha \gamma} = 0 ~,
\label{b2xdiva}
\end{equation}
which means that $f_1 = 0$.
In order to find $f_2(T)$, we will consider a power law dependence and use the fact that $I_2$ has conformal weight equal to 2. Recalling that $\Delta^{\alpha\gamma}$ and $S_{(1)}^{\rho\sigma\mu\nu}$ have conformal weights 2 and 4, respectively, we immediately obtain $f_2=g T^{-8}$, where $g$ is a constant.

So, the final expression for $I_2$ becomes
\begin{equation}
I_2^{\alpha\gamma}= gT^{-8}\Delta^{\alpha \gamma} \xi_{\rho\sigma}\xi^{\rho\sigma} ~.
\label{finI2}
\end{equation}

We have proven that the DTT satisfies the Second Law for arbitrary values of the nonequilibrium tensor $\xi^{\mu\nu}$, provided $I_2$ is given by Eq. (\ref{finI2}). The entropy production is simply (recall Eq. (\ref{D}))
\begin{equation}
S^\mu_{; \mu} = D^{\alpha\gamma\rho\sigma}\xi_{\rho\sigma}\xi_{\alpha\gamma} = \frac{b^2}{\eta T}\xi^{\rho\sigma}\xi_{\rho\sigma}~.
\label{eprod}
\end{equation}

\subsection{Exact hydrodynamic equations}
\label{subeqs}
In this subsection we will obtain the explicit form of the equations of motion of the DTT.

We first turn our attention to the quadratic part of the stress-energy tensor, $\tau_2$. From Eq. (\ref{tau2fromGamma}) we get after some algebra
\begin{equation}
\tau_2^{\mu\nu} = \tilde{c}_1 T^{-4}\bigg(\xi^{\mu\alpha}\xi_\alpha^\nu-\frac{1}{3}\Delta^{\mu\nu}\xi^{\alpha\gamma}\xi_{\alpha\gamma} \bigg) ~,
\label{finaltau2}
\end{equation}
where $\tilde{c}_1=2c_1+c_2$.
We note that $\beta_\mu \tau_2^{\mu\nu} = \beta_\nu \tau_2^{\mu\nu} = 0$ and, of course (we calculated the coefficients for this to happen) $\tau_{2 \mu}^\mu = 0$. The divergence of $\tau_2^{\mu\nu}$ reads,
\begin{equation}
\begin{split}
\tau_{2 ;\nu}^{\mu\nu} &= -4\tilde{c}_1T^{-2}\bigg(\xi^{\mu\alpha}\xi^\nu_\alpha - \frac{1}{3}\Delta^{\mu\nu}\xi^{\alpha\gamma}\xi_{\alpha\gamma} \bigg)\beta^\pi \beta_{\pi ;\nu} \\
&~ -\frac{1}{3}\tilde{c}_1T^{-2}\xi^{\alpha\gamma}\xi_{\alpha\gamma}(\beta^\mu\Delta^{\nu\pi}+\beta^\nu \Delta^{\mu\pi}) \beta_{\pi ;\nu} \\
&~ +\tilde{c}_1T^{-4}\bigg(\xi^{\nu\theta}\xi^\mu_{\theta ;\nu} + \xi^{\mu\theta}\xi^\nu_{\theta ;\nu} \bigg) \\
&~ - \frac{2}{3}\tilde{c}_1 T^{-4}\Delta^{\mu\nu}\xi^{\pi\theta}\xi_{\pi\theta ;\nu} ~.
\end{split}
\label{divtau2fin}
\end{equation}

From Eqs. (\ref{tperf}), (\ref{divtau}) and (\ref{divtau2fin}), the conservation of the complete stress-energy tensor becomes
\begin{equation}
\begin{split}
\tau_{ ;\nu}^{\mu\nu} &= \bigg[ \frac{a}{3}T^{6}(4u^\mu u^\nu + g^{\mu\nu}) \\
&~ - 4\tilde{c}_1T^{-2}\bigg(\xi^{\mu\alpha}\xi^\nu_\alpha - \frac{1}{3}\Delta^{\mu\nu}\xi^{\alpha\gamma}\xi_{\alpha\gamma} \bigg) \bigg] \beta^\pi \beta_{\pi ;\nu} \\
&~ + \bigg(\frac{a}{3}T^{6}- \frac{1}{3}\tilde{c}_1T^{-2}\xi^{\alpha\gamma}\xi_{\alpha\gamma}\bigg) (\beta^\mu\Delta^{\nu\pi}+\beta^\nu \Delta^{\mu\pi}) \beta_{\pi ;\nu} \\
&~ +\tilde{c}_1T^{-4}\bigg(\xi^{\nu\theta}\xi^\mu_{\theta ;\nu} + \xi^{\mu\theta}\xi^\nu_{\theta ;\nu} \bigg) \\
&~ - \frac{2}{3}\tilde{c}_1 T^{-4}\Delta^{\mu\nu}\xi^{\pi\theta}\xi_{\pi\theta ;\nu} + b\xi^{\mu\nu}_{ ;\nu} = 0 ~,
\end{split}
\label{divtaufull}
\end{equation}
where we used that $\rho = aT^{4}$ and $p=\rho/3$.

In the spirit of divergence-type theories, the stress-energy tensor conservation should be supplemented with $A^{\delta\alpha\gamma}_{ ;\delta}=I^{\alpha\gamma}$, which stands on the same footing as the conservation equations \cite{libro,calz98,liu}. We have already obtained $I^{\alpha\gamma}$ in section \ref{secondsec}. Together, they completely describe the space-time evolution of the system (within the hydrodynamic approximation).

In this Section, we completed our first task of finding the generating function $\chi$ and the source tensor $I_2^{\alpha\gamma}$ that describes a conformal fluid in flat space-time. We constructed the DTT by requiring: (i) $\chi$ is quadratic in deviations from equilibrium, represented by the dissipative tensor $\xi$; (ii) the stress-energy tensor derived from $\chi$ is traceless and has the correct conformal weight; (iii) the theory reproduces, at first order, the relativistic Navier-Stokes stress-energy tensor; and (iv) the theory satisfies the Second Law for {\it arbitrary} $\xi$. We have also obtained the equations of motion of the exact theory, which will be used in Section \ref{boost} in the context of Bjorken expansion.

\section{Causality}
\label{caus}

In this section we investigate the causality properties of the DTT constructed above. As noted in Section \ref{app1}, causality is determined solely by the generating function $\chi$ of the theory, and not by the source tensor $I^{\alpha\gamma}$. In order to analyze causality of a DTT, let us define
\begin{equation}
\begin{split}
\zeta^A &= (\beta^\mu,\xi^{\mu\nu})  ~~\textrm{and} \\
M^\mu_{A,B} &= \partial^3 \chi/\partial \beta_\mu \partial \zeta^A \partial \zeta^B  ~,
\end{split}
\end{equation}
being $(A,B)$ collective indices. The DTT is causal (in a set of fluid states near equilibrium) if the vector $M^\mu_{B,C}\delta \zeta^B \delta \zeta^C$ is time-like and future oriented for any displacements $(\delta \zeta^B,\delta \zeta^C)$ from an equilibrium state \cite{calz98,geroch,geraof,liu,marc,marcprd,nagy}.

Since we are interested in proving causality for fluid states near equilibrium, it will be sufficient to deal with
\begin{equation}
M^\mu_{A,B}\bigg|_E = \bigg(\frac{\partial^3 \chi}{\partial \beta_\mu \partial \zeta^A \zeta^B}\bigg)\bigg|_{\zeta^A = \zeta^A_E}
\end{equation}
where
\begin{equation}
\zeta^A_E=(\frac{u^\nu_E}{T_E},\xi^{\nu\delta}=0)
\end{equation}
denotes equilibrium values.

The only non-vanishing terms of $M^\mu_{A,B}\bigg|_E$ are
\begin{equation}
\begin{split}
M^\mu_{\nu,\delta}\bigg|_E &= \bigg(\frac{\partial^3 \chi_{(0)}}{\partial \beta_\mu \partial \beta^\nu \beta^\delta}\bigg) \bigg|_{\zeta^A_E} \\
M^\mu_{\nu\delta,\pi}\bigg|_E &= \bigg(\frac{\partial^3 \chi_{(1)}}{\partial \beta_\mu \partial \beta^\pi \partial \xi^{\nu\delta}}\bigg) \bigg|_{\zeta^A_E} = \bigg(\frac{\partial A^\mu_{E\nu\delta}}{\partial\beta_\pi}\bigg)\bigg|_{\zeta^A_E} ~~ \textrm{and} \\
M^\mu_{\nu\delta,\pi\theta}\bigg|_E &= \bigg(\frac{\partial^3 \chi_{(2)}}{\partial \beta_\mu \partial \xi^{\nu\delta} \partial \xi^{\pi\theta}}\bigg) \bigg|_{\zeta^A_E} = \bigg(\frac{\partial A^\mu_{2\nu\delta}}{\partial\xi^{\pi\theta}}\bigg)\bigg|_{\zeta^A_E}~,
\end{split}
\end{equation}
where $\chi_{(0)}$, $\chi_{(1)}$ and $\chi_{(2)}$ are given by Eqs. (\ref{chi0final}), (\ref{chi1final}) and (\ref{chi2final}), respectively.

Performing the corresponding derivatives, we get
\begin{equation}
\begin{split}
M^\mu_{\nu,\delta}\bigg|_E &= 8aT_E^6\bigg[ (6T_E^2\beta_\nu\beta_\delta+\delta_{\delta\nu})\beta^\mu+\beta_\delta \delta^\mu_\nu +\beta_\nu\delta^\mu_\delta\bigg]_E \\
M^\mu_{\nu\delta,\pi}\bigg|_E &= b\bigg(\delta_{\nu\pi}\delta^\mu_\delta+\delta_{\delta\pi}\delta^\mu_\nu -\frac{1}{2} g^\mu_\pi \delta_{\nu\delta}\bigg) ~~\textrm{and} \\
M^\mu_{\nu\delta,\pi\theta}\bigg|_E &= T_E^{-4}\bigg( 2F^\mu_{\nu\delta\pi\theta} -\frac{1}{2} g^{\rho\sigma}F^\mu_{\nu\delta\rho\sigma}\delta_{\pi\theta}\bigg)_E ~.
\end{split}
\label{Ms}
\end{equation}
where, for brevity, we have defined the tensor
\begin{equation}
\begin{split}
F^\mu_{\nu\delta\rho\sigma} &= c_1\bigg(-6\beta^\mu S_{(1)\nu\delta\rho\sigma} + \frac{1}{3}\beta^\mu\Delta_{\nu\delta}\Delta_{\rho\sigma}+T^2\Delta_{\rho\sigma}\beta_\nu\beta_\delta\beta^\mu \\
&~ +\beta_\nu S_{(1)\delta\rho\sigma}^\mu + \beta_\delta S_{(1)\nu\rho\sigma}^\mu-\frac{2}{3}(\Delta^\mu_\nu \beta_\delta+\beta_\nu\Delta^\mu_\delta)\Delta_{\rho\sigma}\bigg) \\
&~ + 2c_2\Delta^\mu_\sigma\bigg( \Delta_{\rho\nu}\beta_\delta+\beta_\nu\Delta_{\rho\delta}\bigg) ~.
\end{split}
\label{FdeM}
\end{equation}

Since Eqs. (\ref{Ms}) are covariant, we can use any frame to study causality of our DTT. The frame $u^\mu=(1,\vec{0})$ turns out to be very convenient. In this frame we have
\begin{equation}
\delta\beta_\nu=(-t,\vec{w})
\label{deltabeta}
\end{equation}
and
\begin{equation}
\delta \xi^{\mu\nu}=
\begin{bmatrix}
 A & B^1 & B^2 & B^3 \\
B^1  & d_1 &  &  \\
B^2 &  & d_2 &  \\
B^3  &  &  & d_3
\end{bmatrix} ~.
\label{deltaximat}
\end{equation}
In writing the above equations, we have used the general decompositions of a vector $V^\mu$ and a tensor $W^{\mu\nu}$ in time- and space-like parts
\begin{equation}
\begin{split}
V^\mu &= V u^\mu+X^\mu ~~\textrm{and} \\
W^{\mu\nu} &= Au^\mu u^\nu + B^\mu u^\nu + u^\mu C^\nu + E^{\mu\nu}
\end{split}
\end{equation}
with
\begin{equation}
\begin{split}
V &=-u_\mu V^\mu ~,~~ X^\mu = \Delta^\mu_\nu V^\nu \\
A &= W^{\mu\nu}u_\mu u_\nu \\
B^\mu &=-\Delta^\mu_\nu W^{\nu\alpha}u_\alpha \\
C^\mu &=-\Delta^\mu_\nu W^{\alpha\nu}u_\alpha ~~\textrm{and} \\
E^{\mu\nu} &= \Delta^\mu_\alpha \Delta^\nu_\rho W^{\alpha\rho} ~.
\end{split}
\end{equation}
Note that $X^\mu u_\mu = B^\mu u_\mu = C^\mu u_\mu = E^{\mu\nu}u_\mu = E^{\mu\nu}u_\nu= 0$, and that, being real and symmetric, $E^{ij}$ can be diagonalized. In the case of $\delta \xi^{\mu\nu}$ we have put $\delta \xi^{i j}=\textrm{diag}(d_1,d_2,d_3)$. Since $\delta \xi^{\mu\nu}$ should remain traceless and symmetric (as $\xi^{\mu\nu}$), we have $B^\mu=C^\mu=(0,\vec{B})$ and $d_1+d_2+d_3=A$.

In the frame $u^\mu=(1,\vec{0})$ we have
\begin{equation}
M^\mu_{\nu,\delta}\bigg|_E = 8aT_E^5 \bigg[ \delta^\mu_0 (6\delta_{\nu 0}\delta_{\delta 0}+\delta_{\delta \nu})-\delta^\mu_\nu \delta_{\delta 0}-\delta^\mu_\delta \delta_{\nu 0}\bigg]
\label{M1enframe}
\end{equation}
and
\begin{equation}
\begin{split}
T_E^5 M^\mu_{\nu\delta,\pi\theta}\bigg|_E &= 2c_1 \delta^\mu_0\bigg(-3\delta_{\nu i} \delta_{\pi i}\delta_{\delta j}\delta_{\theta j}-3\delta_{\nu i} \delta_{\theta i}\delta_{\delta j}\delta_{\pi j} \\
& ~+\frac{7}{3}\delta_{\nu i} \delta_{\delta i}\delta_{\pi j}\delta_{\theta j} +\delta_{\pi i} \delta_{\theta i}\delta_{\nu 0}\delta_{\delta 0}\bigg) \\
& ~-c_1 \delta^\mu_i\delta_{\nu 0}\bigg(\delta_{\pi i}\delta_{\delta j}\delta_{\theta j} - 2 \delta_{\delta i}\delta_{\pi j}\delta_{\theta j}\bigg) \\
& ~-c_1 \delta^\mu_i \delta_{\delta 0}\bigg(\delta_{\pi i}\delta_{\nu j}\delta_{\theta j} - 2 \delta_{\nu i}\delta_{\pi j}\delta_{\theta j}\bigg) \\
& ~-(c_1+4c_2)\delta^\mu_i \bigg(\delta_{\delta 0}\delta_{\theta i}\delta_{\nu j}\delta_{\pi j} + \delta_{\nu 0}\delta_{\delta i}\delta_{\pi j}\delta_{\theta j}\bigg) \\
& ~-\frac{c_1}{2}\delta^\mu_0\delta_{\pi\theta}(\delta_{\nu i}\delta_{\delta i}-3\delta_{\nu 0}\delta_{\delta 0}) \\
& ~+(c_2-c_1)\delta^\mu_i(\delta_{\nu i}\delta_{\delta 0}+\delta_{\nu 0}\delta_{\delta i})\delta_{\pi\theta} ~.
\end{split}
\label{M3enframe}
\end{equation}
Note that $M^\mu_{\nu\delta,\pi}$ is frame invariant (see Eq. (\ref{Ms})).

Putting
\begin{equation}
\begin{split}
r^\mu &= M^\mu_{\nu,\delta}\bigg|_E \delta \beta^\nu \delta \beta^\delta \\
z^\mu &= M^\mu_{\nu\delta,\pi}\bigg|_E \delta\xi^{\nu\delta} \delta\beta^\pi  ~~\textrm{and} \\
s^\mu &= M^\mu_{\nu\delta,\pi\theta}\bigg|_E \delta \xi^{\nu\delta} \delta \xi^{\pi\theta}
\end{split}
\end{equation}
we get, from Eqs. (\ref{deltabeta}), (\ref{deltaximat}), (\ref{M1enframe}) and (\ref{M3enframe}),
\begin{equation}
\begin{split}
r^\mu &= 8aT_E^5\bigg(5(\delta\beta^0)^2+\sum_{i=1}^3 (\delta \beta^i)^2,-2\delta\beta^0 \delta \vec{\beta}\bigg) \\
&= 8aT_E^5\bigg(5t^2+\sum_{i=1}^3 w_i^2 , -2t \vec{w} \bigg) ~,
\end{split}
\end{equation}
\begin{equation}
z^\mu= b\delta\beta_\nu \delta \xi^{\mu\nu} = b\bigg( -tA+\vec{B}\cdot \vec{w},-tB^j+d_jw^j \bigg)
\label{zmu}
\end{equation}
and
\begin{equation}
s^\mu = T_E^{-5}\bigg(c_1[12G+\frac{26}{3}A^2], -[c_1(3d_j + A)+4c_2 d_j]B^j  \bigg) ~,
\label{smu}
\end{equation}
where $G\equiv \sum_i (d_i)^2$, $j=(1,2,3)$ and no sum is implied in the spatial part. So, we must now see whether $y^\mu=r^\mu+z^\mu+s^\mu$ is time-like and future oriented, {\it for arbitrary values of the fluctuations}. It is clear that $r^\mu$ poses no problem; it is time-like and future oriented if $a>0$. This was expected since $r^\mu$ corresponds to a perfect fluid.

Before analyzing the vectors $z^\mu$ and $s^\mu$, it will be convenient to determine the constraints that the transversality of $\xi^{\mu\nu}$ imposes on them. We have found before that, in order for the DTT to be consistent, the nonequilibrium tensor had to be transverse. From a physical point of view, this meant that the heat flow and the bulk viscosity remain zero when the conformal fluid departs from equilibrium. This is a sensible result since we want the theory to remain conformal even in the presence of dissipation. Therefore, the condition we must impose is
\begin{equation}
\beta_\mu (\xi^{\mu\nu}|_E+\delta \xi^{\mu \nu}) = (\beta_\mu|_E + \delta \beta_\mu)\delta \xi^{\mu \nu} =0
\label{fulltrans}
\end{equation}
where we have used that $\xi^{\mu\nu}|_E=0$. Using Eqs. (\ref{deltabeta}) and (\ref{deltaximat}) we get
\begin{equation}
\beta_\mu\delta \xi^{\mu \nu} = \bigg(-\tilde{t}A + \vec{w}\cdot \vec{B}, -\tilde{t}B^j+d_j w^j \bigg)
\label{explicittrans}
\end{equation}
with $\tilde{t}=(t+T_E^{-1})$, being $T_E$ the equilibrium temperature of the fluid. Requiring $\delta \xi^{\mu \nu}$ to be transverse we obtain
\begin{equation}
\begin{split}
\tilde{t}A &= \vec{w}\cdot \vec{B} ~~\textrm{and} \\
\tilde{t}B^j &= d_jw^j  ~~ \textrm{(no sum in $j$)}
\end{split}
\label{transsystem}
\end{equation}
or
\begin{equation}
A=\frac{1}{\tilde{t}^2}\sum_{j=1}^3 d_jw_j^2 ~.
\end{equation}

Using these results obtained from the transversality of $\xi^{\mu\nu}$, we can rewrite the vector $z^\mu$ in a simple way:
\begin{equation}
z^\mu=\frac{b}{T_E}\bigg( A,\vec{B} \bigg) ~.
\label{simplez}
\end{equation}
Therefore, putting $l^\mu=z^\mu+s^\mu$ we have
\begin{equation}
\begin{split}
l^0 &= \frac{b}{T_E}A + \frac{c_1}{T_E^5}(12G+\frac{26}{3}A^2) ~~\textrm{and} \\
l^i &= \bigg( \frac{b}{T_E}-T_E^{-5}[c_1(3d_i+A)+4c_2 d_i] \bigg) B^i ~.
\end{split}
\end{equation}
In order to prove that $(l^0)^2 > (l^i)^2$ (i.e. that $l^\mu$ is time-like) it is convenient to reexpress $l^\mu$ in tensorial notation. From Eq. (\ref{transsystem}) we get
\begin{equation}
\begin{split}
\vec{B} &= \frac{1}{\tilde{t}} ({\mathbf d} \cdot \vec{w}) ~~~~\textrm{and} \\
\textrm{Tr}({\mathbf d}) &= \frac{1}{\tilde{t}^2}(\vec{w} \cdot {\mathbf d} \cdot \vec{w}) ~, 
\end{split}
\end{equation}
where ${\mathbf d}=d_{ij}$ is the spatial part of $\delta \xi_{\mu\nu}$, which we do not assume to be diagonal. Using these relations, $l^\mu$ becomes
\begin{equation}
\begin{split}
l^0 &= \frac{b}{\tilde{t}^2 T_E} \vec{w} \cdot {\mathbf d} \cdot \vec{w} + \frac{26 c_1}{3 \tilde{t}^4 T_E^5}(\vec{w} \cdot {\mathbf d} \cdot \vec{w})^2 +\frac{12 c_1}{T_E^5}{\mathbf d}:{\mathbf d} ~~\textrm{and} \\
\vec{l} &= \frac{b}{\tilde{t} T_E}{\mathbf d} \cdot \vec{w} - \frac{c_1}{\tilde{t}^3 T_E^5}(\vec{w} \cdot {\mathbf d} \cdot \vec{w})({\mathbf d} \cdot \vec{w}) \\
&~ -\frac{(3c_1+4c_2)}{\tilde{t} T_E^5}({\mathbf d}^2 \cdot \vec{w}) ~, 
\end{split}
\label{newlmu}
\end{equation}
where ${\mathbf d}:{\mathbf d}$ stands for $\textrm{Tr} ([{\mathbf d}]^2)$. It is clear that if $\vec{w}=0$, $l^\mu$ is trivially time-like and future-oriented, provided $c_1>0$, and therefore the theory is causal in this case. By continuity, the DTT will remain causal provided $\vec{w}$ is not too large. One can actually quantify this (at lowest order) by keeping linear terms in $\vec{w}$ and requiring that $(l^0)^2 > \vec{l}\cdot \vec{l}$, but the resulting expression is not too illuminating.

\section{Adiabatic expansion in velocity gradients}
\label{eqsmotion}

In this section, we set up a consistent adiabatic expansion of the DTT to compare with previous approaches based on conformal invariants, first put forward in Refs. \cite{sonhydro,bat}. As stated in the Introduction, we will limit ourselves to Minkowski space-time.

For a conformal fluid in flat space-time, the dissipative part of the stress-energy tensor complete at second-order {\it in velocity gradients} can be written as \cite{bat,logan,sonhydro,romentropy09}
\begin{equation}
\begin{split}
\tau^{\mu\nu}_{c.i.} &= -\eta \sigma^{\mu\nu} +\eta \tau_\pi \bigg( S_{(1)}^{\mu\nu\rho\sigma}D\sigma_{\rho\sigma}
+\frac{1}{3}\sigma^{\mu\nu}(u^\delta_{ ;\delta})\bigg) \\
&~ +\frac{\lambda_1}{\eta^2} S_{(1)}^{\mu\nu\rho\sigma} \sigma_\rho^\lambda \sigma_{\sigma\lambda} + \frac{\lambda_2}{\eta} S_{(1)}^{\mu\nu\rho\sigma}\sigma_\rho^\lambda \Omega_{\rho\lambda} \\
&~ + \lambda_3 S_{(1)}^{\mu\nu\rho\sigma}\Omega_\rho^\lambda \Omega_{\rho\lambda}
\end{split}
\label{tauson}
\end{equation}
where the subscript $c.i.$ is a remainder that this form of $\tau_2$ is constructed from conformal invariants (as explained in detail in Refs. \cite{bat,logan,sonhydro,romentropy09}).
$D=u_\mu\partial^\mu$ is the convective time derivative, $(\tau_\pi,\lambda_i)$ are second-order transport coefficients, and $\Omega_{\rho\lambda}$ is the vorticity. As already mentioned, this expression for $\tau_2$ represents an extension of Israel-Stewart entropy-wise approach.

We will now show that, for the case $\lambda_{2,3}=0$, $\tau^{\mu\nu}_{c.i.}$ can be obtained from a consistent adiabatic expansion (at second-order in velocity gradients) of the exact hydrodynamic equations. We start by requiring that $\tau_2^{\mu\nu}$ calculated from $\chi$ be equal to $\tau^{\mu\nu}_{c.i.}$ calculated from second-order (in velocity gradients) conformal invariants. We have
\begin{equation}
\begin{split}
b\xi^{\mu\nu}+H^{\mu\nu\alpha\gamma\rho\sigma}\xi_{\alpha\gamma}\xi_{\rho\sigma}&= -\eta \sigma^{\mu\nu} +\eta \tau_\pi \bigg( S_{(1)}^{\mu\nu\rho\sigma}D\sigma_{\rho\sigma} \\
&~ +\frac{1}{3}\sigma^{\mu\nu}(u^\delta_{ ;\delta})\bigg) +\frac{\lambda_1}{\eta^2} S_{(1)}^{\mu\nu\rho\sigma} \sigma_\rho^\lambda \sigma_{\sigma\lambda} \\
&~ + \frac{\lambda_2}{\eta} S_{(1)}^{\mu\nu\rho\sigma}\sigma_\rho^\lambda \Omega_{\rho\lambda} +
\lambda_3 S_{(1)}^{\mu\nu\rho\sigma}\Omega_\rho^\lambda \Omega_{\rho\lambda}
\end{split}
\label{sontau}
\end{equation}
where we have put (see Eq. (\ref{finaltau2}))
\begin{equation}
\begin{split}
H^{\mu\nu\alpha\gamma\rho\sigma} &= \Gamma^{\mu\nu} \sum_i c_i S_{(i)}^{\alpha\gamma\rho\sigma} = \tilde{c}_1 T^{-4}\bigg( \frac{1}{4}(\delta^{\alpha\mu}S_{(1)}^{\nu\gamma\rho\sigma} \\
&~ + \delta^{\gamma\mu}S_{(1)}^{\alpha\nu\rho\sigma} + \delta^{\rho\mu}S_{(1)}^{\alpha\gamma\nu\sigma}
+ \delta^{\sigma\mu}S_{(1)}^{\alpha\gamma\rho\nu}) \\
&~ - \frac{1}{3}\Delta^{\mu\nu}S_{(1)}^{\alpha\gamma\rho\sigma} \bigg) ~.
\end{split}
\label{exH}
\end{equation}

Putting
\begin{equation}
\xi^{\mu\nu}=\xi_{(1)}^{\mu\nu}+\xi_{(2)}^{\mu\nu} = -\frac{\eta}{b}\sigma^{\mu\nu}+\xi_{(2)}^{\mu\nu}
\end{equation}
in Eq. (\ref{sontau}), and retaining terms up to second order we get
\begin{equation}
\begin{split}
\xi^{\mu\nu}_{(2)} &= \eta\tau_\pi \bigg( S_{(1)}^{\mu\nu\rho\sigma}D\sigma_{\rho\sigma}+\frac{1}{3}\sigma^{\mu\nu}(u^\delta_{ ;\delta}) \bigg) \\
&~ +\frac{\lambda_1}{\eta^2} S_{(1)}^{\mu\nu\rho\sigma} \sigma_\rho^\lambda \sigma_{\sigma\lambda} + \frac{\lambda_2}{\eta} S_{(1)}^{\mu\nu\rho\sigma}\sigma_\rho^\lambda \Omega_{\rho\lambda} \\
&~ + \lambda_3 S_{(1)}^{\mu\nu\rho\sigma}\Omega_\rho^\lambda \Omega_{\rho\lambda} -\frac{\eta^2}{b^2}H^{\mu\nu\alpha\gamma\rho\sigma}\sigma_{\alpha\gamma}\sigma_{\rho\sigma} ~.
\end{split}
\label{xi2exp}
\end{equation}
Using Eq. (\ref{exH}), we can rewrite the last equation more explicitly
\begin{equation}
\begin{split}
\xi^{\mu\nu}_{(2)} &= \eta\tau_\pi \bigg(~^<D\sigma^{\mu\nu>}+\frac{1}{3}\sigma^{\mu\nu}(u^\delta_{ ;\delta}) \bigg) \\
&~ +\bigg( \frac{\lambda_1}{\eta^2} -\frac{\eta^2\tilde{c}_1 T^{-4}}{b^2}\bigg)\sigma^{<\mu\lambda} \sigma_{\lambda}^{\nu>} +\frac{\eta^2\tilde{c}_1 T^{-4}}{3b^2}\Delta^{\mu\nu}\sigma^{\rho\sigma}\sigma_{\rho\sigma}\\
&~ + \frac{\lambda_2}{\eta} \sigma^{<\mu\lambda} \Omega^{\nu>}_\lambda 
+ \lambda_3 \Omega^{<\mu\lambda} \Omega_{\lambda}^{\nu>}  ~,
\end{split}
\label{xi2more}
\end{equation}
where we introduced $<\ldots>$ to denote the spatial, symmetric and traceless projection of a tensor:
\begin{equation}
B^{<\mu\nu>}=S_{(1)}^{\mu\nu\alpha\gamma}A_{\alpha\gamma} ~.
\end{equation}

At second-order in velocity gradients, the equation $A_{ ;\delta}^{\delta\alpha\gamma}=I^{\alpha\gamma}$ reads
\begin{equation}
\begin{split}
& \bigg(\frac{\partial A_E^{\delta\alpha\gamma}}{\delta \beta_\pi} + \frac{\partial G^{\delta\alpha\gamma\rho\sigma}}{\partial \beta_\pi}\xi^{(1)}_{\rho\sigma}\bigg) \beta_{\pi;\delta}+G^{\delta\alpha\gamma\rho\sigma}\frac{\partial \xi_{\rho\sigma}}{\partial \xi_{\pi\theta}}\xi^{(1)}_{\pi\theta;\delta}\\
&~ = -D^{\alpha\gamma\rho\sigma}(\xi^{(1)}_{\rho\sigma}+\xi^{(2)}_{\rho\sigma})+ gT^{-8}\Delta^{\alpha\gamma}\xi_{(1)}^{\rho\sigma}\xi^{(1)}_{\rho\sigma} ~,
\end{split}
\label{eomexpans}
\end{equation}
where we have used Eqs. (\ref{symbdiva}) and (\ref{finI2}).
Explicitly, we have
\begin{equation}
\begin{split}
D^{\alpha\gamma\rho\sigma}\xi^{(2)}_{\rho\sigma} &= \xi_{(2)}^{\alpha\gamma} = -\frac{\partial G^{\delta\alpha\gamma\rho\sigma}}{\partial \beta_\pi}\xi^{(1)}_{\rho\sigma} \beta_{\pi;\delta}\\
&~ -G^{\delta\alpha\gamma\rho\sigma}\frac{\partial \xi_{\rho\sigma}}{\partial \xi_{\pi\theta}}\xi^{(1)}_{\pi\theta;\delta}
+ gT^{-8}\Delta^{\alpha\gamma}\xi_{(1)}^{\rho\sigma}\xi^{(1)}_{\rho\sigma} ~.
\label{xi2dos}
\end{split}
\end{equation}

The crucial point is that, in order for $\tau^{\mu\nu}_{c.i.}$ to be derivable from the DTT, both expressions for $\xi^{\mu\nu}_{(2)}$, given in Eqs. (\ref{xi2exp}) and (\ref{xi2dos}), should coincide. We see immediately that (actually, this equation holds for the exact $\xi_{\rho\sigma}$)
\begin{equation}
G^{\delta\alpha\gamma\rho\sigma}\frac{\partial \xi_{\rho\sigma}}{\partial \xi_{\pi\theta}}\xi_{\pi\theta;\delta}^{(1)} = G^{\delta\alpha\gamma\rho\sigma}\xi_{\rho\sigma;\delta}^{(1)} ~.
\label{molesto}
\end{equation}
Using Eq. (\ref{diva2}), we can rewrite Eq. (\ref{xi2dos}) as
\begin{equation}
\begin{split}
\xi_{(2)}^{\alpha\gamma} &= 12c_1T^{-4} (4u^\pi u^\delta +\delta^{\pi\delta})\xi^{\alpha\gamma}_{(1)}\beta_{\pi;\delta} \\
&~ +2(2c_1+c_2)T^{-4}\bigg(\xi^{\delta\gamma}_{(1)}\beta^\alpha_{ ;\delta}+\xi^{\delta\alpha}_{(1)}\beta^\gamma_{ ;\delta}\bigg) \\
&~ -12c_1 T^{-5}u^\delta S_{(1)}^{\alpha\gamma\rho\sigma} \xi^{(1)}_{\rho\sigma;\delta} +gT^{-8}\Delta^{\alpha\gamma}\xi_{(1)}^{\rho\sigma} \xi^{(1)}_{\rho\sigma}~.
\end{split}
\label{expandedxi2}
\end{equation}
The third term becomes
\begin{equation}
-12c_1 T^{-5}S_{(1)}^{\alpha\gamma\rho\sigma} D \xi^{(1)}_{\rho\sigma} = \frac{12 \eta c_1}{b}T^{-5} ~^<D\sigma^{\alpha\gamma>} ~,
\end{equation}
which reproduces the first term of Eq. (\ref{xi2more}) if $\tau_\pi=12c_1/(bT^5)$. Using  
that
\begin{equation}
u^\delta_{ ;\delta} = -3 D\ln T ~,
\label{lntdivu}
\end{equation}
it can be seen that the first term of Eq. (\ref{expandedxi2}) reproduces the second term of Eq. (\ref{xi2more}). The second term of Eq. (\ref{expandedxi2}) reproduces the third and fourth terms of Eq. (\ref{xi2more}), provided 
\begin{equation}
\lambda_1 = \frac{\eta^3}{bT^5}\bigg[-4 + \frac{\eta}{b}\bigg](2c_1+c_2) ~.
\end{equation}
The last term of (\ref{expandedxi2}) reproduces the fifth term of Eq. (\ref{xi2more}) if $2c_1+c_2=3gT^{-4}\eta^2$.
However, it is not possible to reproduce, from Eq. (\ref{expandedxi2}), the vorticity terms of Eq. (\ref{xi2more}). So, we conclude that the DTT we have constructed is limited to the case $\lambda_{2,3}=0$. We note that this is not a serious restriction on the application of the DTT to heavy-ion collisions (see especially Ref. \cite{luzum}).

We have proven that (for $\lambda_{2,3}=0$) $\tau^{\mu\nu}_{c.i.}$, as given by Eq. (\ref{tauson}), can be obtained from a consistent adiabatic expansion (at second-order in velocity gradients) of the exact divergence-type theory we have developed. This is one of the most important results of this work.

We have already proven that the DTT satisfies the Second Law. It is clear that its adiabatic expansion satisfies it too. It is interesting to remark that, when expanding the entropy production given in Eq. (\ref{eprod}), i.e. when putting $\xi_{\alpha\gamma}=\xi^{(1)}_{\alpha\gamma}+\xi^{(2)}_{\alpha\gamma}$, terms up to fourth order in velocity gradients arise. This agrees with the entropy production form calculated by Loganayagam in Ref. \cite{logan}, based on the developments of Refs. \cite{sonhydro,bat}. Dropping fourth-order terms in the entropy production, although it may be justified under some circumstances, actually spoils the consistency of the adiabatic expansion (see Ref. \cite{romentropy09} for interesting discussions on higher order terms in the entropy production).

\section{Boost invariant flow}
\label{boost}

We will now obtain the equations of motion of the DTT for the case of Bjorken flow \cite{bj} (see also Refs. \cite{sonhydro,muronga04,rom09,libro}), which, besides of being much more simple than general flow, is a successful toy model of heavy-ion collisions in the mid-rapidity region. The comparison between the equations of the DTT and the second-order ones for the case of boost invariant flow is interesting because it clearly shows the difference between the exact and truncated equations, in a relatively simple situation. In the last part of this section, we compare the numerical solution to the exact and truncated equations.

The motion in the Bjorken flow is a 1D
expansion, along an axis which we choose to be $z$, with local velocity equal to $z/t$. It is convenient to choose comoving coordinates (Milne coordinates), proper time $\tau$ and rapidity $\psi$, given by
\begin{equation}
\tau=\sqrt{t^2-z^2} ~~ \textrm{and} ~~
\psi=\textrm{arctanh}(z/t) ~.
\end{equation}
The advantage of using these coordinates is that each element is at rest: $(u^\tau , u^{\perp}, u^\psi) =
(1, 0, 0)$. Although the velocity vector is constant, the dynamics is nontrivial because not every Christoffel symbol is zero. The metric tensor is
\begin{equation}
g_{\mu\nu}=\textrm{diag}(g_{\tau\tau},g_{xx},g_{yy},g_{\psi\psi})=\textrm{diag}(1,-1,-1,-\tau^2) 
\end{equation}
where $(x,y)$ denote transverse directions, so we have
\begin{equation}
\begin{split}
D\equiv u_\mu\partial^\mu &\rightarrow \partial_\tau ~~ \textrm{and} \\
\partial_\mu u^\mu &\rightarrow \frac{1}{\tau} ~.
\end{split}
\label{dersinbjorken}
\end{equation}
The only nonvanishing component of the dissipative part of the stress-energy tensor is the $(\psi,\psi)$ component. Note that the motion is irrotational, and that the energy density and the dissipative part of the stress-energy tensor only depend on proper time (i.e. are independent of rapidity). The only conservation equation that is nontrivial for Bjorken flow is the energy equation, i.e. $u_\mu T^{\mu\nu}_{ ;\nu}=0$, where $T^{\mu\nu}$ is the complete stress-energy tensor.

\subsection{Second-order theory}

At second order in velocity gradients, the hydrodynamic equations for Bjorken flow are (see Refs. \cite{sonhydro,luzum,rom09} for detailed discussions)
\begin{equation}
\begin{split}
\partial_\tau \rho &= -\frac{\rho+p}{\tau}+\frac{\Pi^{\psi}_\psi}{\tau} ~~~~~\textrm{with}  \\
\partial_\tau \Pi^{\psi}_\psi &= -\frac{\Pi^{\psi}_\psi}{\tau_\pi} +\frac{4\eta}{3\tau_\pi \tau} -\frac{4}{3\tau}\Pi^{\psi}_\psi - \frac{\lambda_1}{2\tau_\pi \eta^2}[\Pi^\psi_\psi]^2 ~,
\end{split}
\label{bjsec}
\end{equation}
where, in the notation used here,
\begin{equation}
\Pi^{\mu\nu} \equiv \tau_1^{\mu\nu}+\tau_2^{\mu\nu} ~.
\label{pitaus}
\end{equation}
Actually, the differential equation for $\Pi^{\psi}_\psi$ showed in Eq. (\ref{bjsec}) is exact up to terms which are second-order in velocity gradients. It is obtained by replacing $\sigma^{\mu\nu}$ by $\Pi^{\mu\nu}$ in the gradient expansion of the latter \cite{sonhydro} (see Eq. (\ref{tauson})).

For a conformal perfect fluid in d=4, $\rho(\tau)=C\tau^{-4/3}$, where $C$ is a constant. Due to conformal invariance, the viscosity and the second-order transport coefficients must scale as follows:
\begin{equation}
\eta=C\eta_0\bigg(\frac{\rho}{C} \bigg)^{3/4} ~~~ \tau_\pi = \tau^0_\pi \bigg(\frac{\rho}{C} \bigg)^{-1/4} ~~~ \lambda_1 = C\lambda^0_1 \bigg(\frac{\rho}{C} \bigg)^{1/2} ~,
\label{scalecoeff}
\end{equation}
where $\eta_0$, $\tau^0_\pi$ and $\lambda^0_1$ are constants. 

Note that the Navier-Stokes equations are recovered formally by setting $\tau_\pi, \lambda_1 \rightarrow 0$, whereby
\begin{equation}
\Pi^\psi_\psi \bigg|_1 = \frac{4\eta}{3 \tau} ~.
\label{nsaux}
\end{equation}

\subsection{Divergence-type theory}

Projection of Eq. (\ref{divtaufull}) onto $u_\mu$ leads to
\begin{equation}
\begin{split}
D\rho &= -\bigg(\rho+p +\frac{1}{3}\tilde{c}_1 T^{-4}\xi^{\alpha\gamma}\xi_{\alpha\gamma} \bigg)\nabla_\mu u^\mu \\
&~ + b\xi^{\mu\nu}\sigma_{\mu\nu}+\tilde{c}_1 T^{-4}\xi^{\mu\alpha}\xi^\nu_\alpha \sigma_{\mu\nu} ~.
\end{split}
\label{eneexact}
\end{equation}

For Bjorken flow, energy conservation reads
\begin{equation}
\partial_\tau \rho = -\frac{1}{\tau}\bigg(\rho+p +\frac{2}{3}(2c_1+c_2) T^{-4} [\xi^{\psi}_\psi]^2 \bigg) \\
+ \frac{b}{\tau}\xi^{\psi}_\psi
\label{bj1exact}
\end{equation}
while the equation $A_{ ;\delta}^{\delta\alpha\gamma}=I^{\alpha\gamma}$ becomes
\begin{equation}
\begin{split}
& \frac{12c_1T^{-5}}{\tau^2} \partial_\tau \xi^{\psi}_\psi -\frac{4b}{3T\tau} +8c_1T^{-5}\frac{\xi^{\psi}_\psi}{\taṻ̱̰̃́̀̌̄^3} \\
&= -\frac{b^2}{\eta T}\xi^{\psi}_\psi + 3g T^{-8}[\xi^{\psi}_\psi]^2 ~,
\end{split}
\label{bj2exact}
\end{equation}
where we made use of Eq. (\ref{lntdivu}).

The DTT as well as the second-order theory reduce to Eckart's theory when retaining first-order velocity gradients. Therefore, it is clear that the hydrodynamic equations of both theories must coincide in that limit (of course, this statement is valid for general flow, but we will discuss Bjorken flow only). Noticing that at first order in gradients we can write $\Pi^{\mu\nu}=b\xi_{(1)}^{\mu\nu}$ and $(c_1,c_2,g,\tau_\pi,\lambda_1) = 0$, we immediately see that the hydrodynamic equations of the DTT and the  second-order theory (Eqs. (\ref{bj1exact},\ref{bj2exact}) and
Eqs. (\ref{bjsec}), respectively) coincide. From Eq. (\ref{bj2exact}) we recover the Navier-Stokes limit given in Eq. (\ref{nsaux}). 

The comparison between the equations of both theories beyond first order in velocity gradients becomes quite complicated, because, being $\tau_2^{\mu\nu}$ quadratic in $\xi^{\mu\nu}$, $\Pi^{\mu\nu}$ and $\xi^{\mu\nu}$ are not linearly related anymore (see Eq. (\ref{pitaus})). In order to carry out the this comparison, we solve both sets of differential equations numerically in the next subsection. Before doing that, it is convenient to reexpress (using the results of the previous section) the equations of the DTT in terms of $(\eta,\tau_\pi,\lambda_1)$ instead of $(b,c_1,c_2)$. Without loss of generality, we can fix $b=\eta$ (this means $\xi^{\mu\nu}_{(1)}=-\sigma^{\mu\nu}$), whereby 
\begin{equation}
c_2=-\frac{T^5}{3}\bigg(\frac{\eta\tau_\pi}{2}+\frac{\lambda_1}{\eta^2} \bigg) ~~\textrm{and} ~~ c_1=\frac{\eta T^5 \tau_\pi}{12} ~. 
\end{equation}
Note that $g$ is completely specified once $c_1$ and $c_2$ are known:
\begin{equation}
g= -\frac{\lambda_1 T^9}{9\eta^4} ~.
\end{equation}
Equations (\ref{bj1exact}) and (\ref{bj2exact}) then read
\begin{equation}
\partial_\tau \rho = -\frac{1}{\tau}\bigg(\rho+p -T F_1 [\xi^{\psi}_\psi]^2 \bigg) \\
+ \frac{\eta}{\tau}\xi^{\psi}_\psi
\label{bj1exactnew}
\end{equation}
with 
\begin{equation}
F_1=\frac{2\lambda_1}{9\eta^2} ~,
\end{equation}
and
\begin{equation}
\partial_\tau \xi^\psi_\psi = E_1 +E_2\xi^\psi_\psi + E_3 [\xi^\psi_\psi]^2 
\label{bj2exactnew}
\end{equation}
with
\begin{equation}
\begin{split}
E_1 &= \frac{4\tau }{3T \tau_\pi} \\ 
E_2 &= -\bigg( \frac{2}{3\tau}+\frac{\tau^2}{T\tau_\pi}\bigg) \\
E_3 &= -\frac{\lambda_1 T \tau^2}{3\tau_\pi \eta^5} ~.
\end{split}
\end{equation}
The dissipative part of the stress-energy tensor in the DTT is constructed from the solution to Eq. (\ref{bj2exactnew}). We have
\begin{equation}
\tau_{1\psi}^\psi + \tau_{2\psi}^\psi= \eta \xi_\psi^\psi +F_1 T[\xi^\psi_\psi]^2 ~.
\end{equation}

\subsection{Comparison of numerical solutions}

In this section, we compare the solutions to the hydrodynamic equations of the DTT, second-order and Navier-Stokes theories. We focus on the inverse Reynold's number 
\begin{equation}
R^{-1}=\frac{\Pi^\psi_\psi}{\rho+p} ~,
\end{equation}
and on the pressure isotropy 
\begin{equation}
\frac{P_L}{P_T} = \frac{p-\Pi^\psi_\psi}{p+\Pi^\psi_\psi/2} ~.
\end{equation}
These two quantities are relevant parameters to characterize the hydrodynamic evolution (see, for instance, Refs. \cite{houv,muronga04,baier06,ael,rischke}). Ideal fluids are characterized by $R^{-1}=0$ and $P_L/P_T=1$. Note that, as already mentioned, in the DTT we have $\Pi^\psi_\psi=\tau_{1\psi}^\psi + \tau_{2\psi}^\psi$. 
When solving the hydrodynamic equations, one must bear in mind that the transport coefficients are functions of the energy density $\rho$, as given by Eq. (\ref{scalecoeff}). In particular, we will focus on the strongly-coupled SYM plasma, for which we have \cite{sonhydro}
\begin{equation}
\tau_\pi = 2(2-\ln 2)\frac{\eta}{sT} ~~~\textrm{and} ~~~ \lambda_1=\frac{\eta}{2\pi T} ~,
\end{equation}
where $s$ is the entropy density.

In the following, we present the results for two relevant  values of $\eta/s$. This value is modified by changing the value of $\eta_0$. We consider $\eta/s=0.09$, which is very close to the lower bound imposed by the AdS/CFT correspondence ($\eta/s\ge 1/4\pi$), and $\eta/s=0.375$, which is close to the upper bound for the quark-gluon plasma found by comparing dissipative hydrodynamics to elliptic flow measurements ($\eta/s\le 0.5$) \cite{luzum}.  
As initial conditions, we set $\Pi^\psi_\psi(\tau_0)=0$, $\tau_0=0.5$ fm/c and $\rho(\tau_0)=10$ GeV/fm$^3$ in all calculations. 

\begin{figure}
\includegraphics{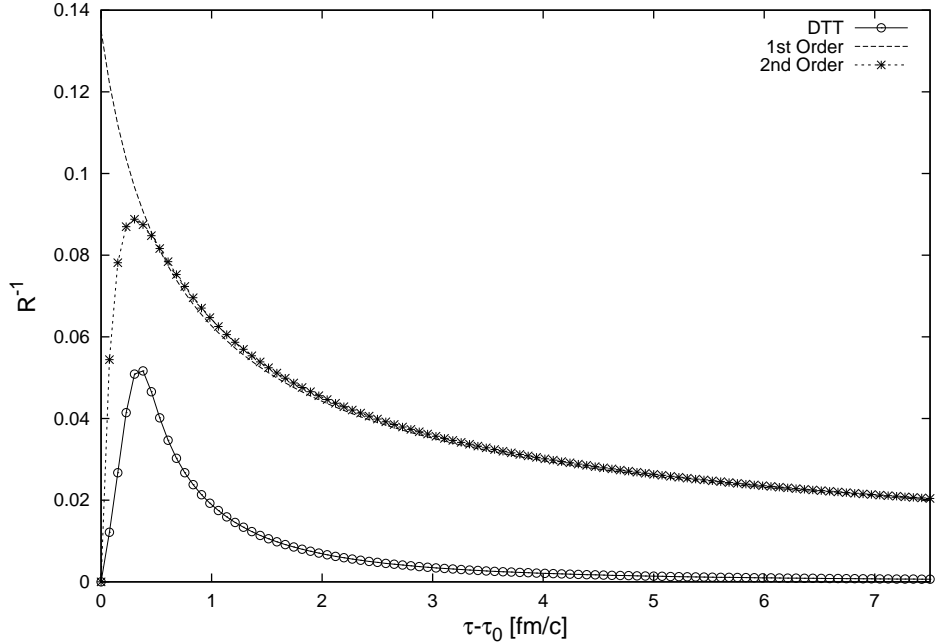}
\vspace{1cm}
\caption{Inverse Reynold's number $R^{-1}$ as a function of proper time, for the DTT, second-order and Navier-Stokes theories with $\eta/s=0.09$.}
\label{compR1}
\end{figure}
\begin{figure}
\includegraphics{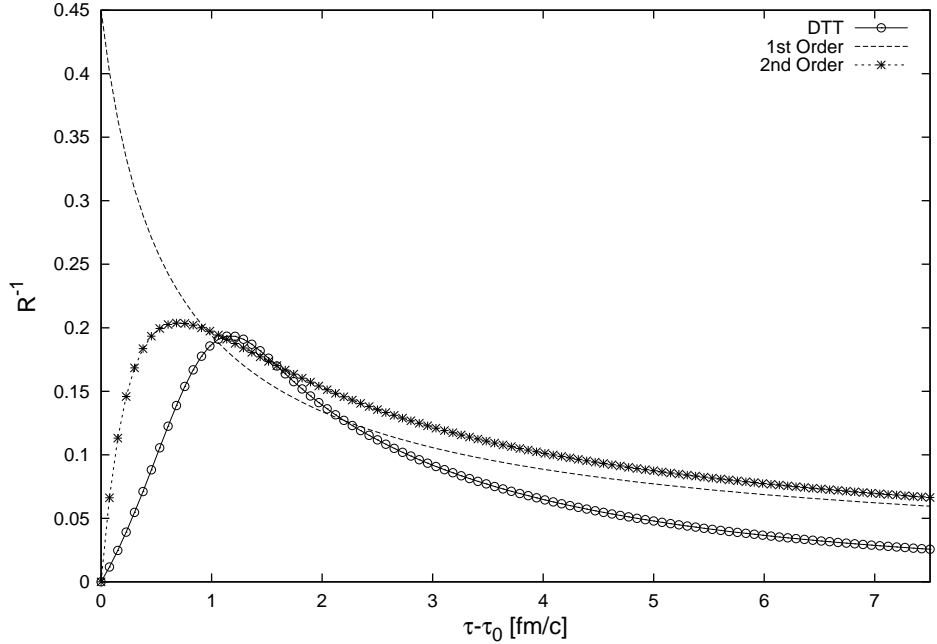}
\vspace{1cm}
\caption{Inverse Reynold's number $R^{-1}$ as a function of proper time, for the DTT, second-order and Navier-Stokes theories with $\eta/s=0.375$.}
\label{compR2}
\end{figure}

In Figure \ref{compR1} we compare the evolution of the inverse Reynold's number with proper time for the DTT, the second-order and Navier-Stokes theories with $\eta/s=0.09$. The most important feature is that the DTT shows a faster approach to ideal hydrodynamics.  
Figure \ref{compR2} shows the same comparison but for $\eta/s=0.375$. As in the previous case, the DTT shows a faster approach to the ideal fluid behaviour. 

\begin{figure}
\includegraphics{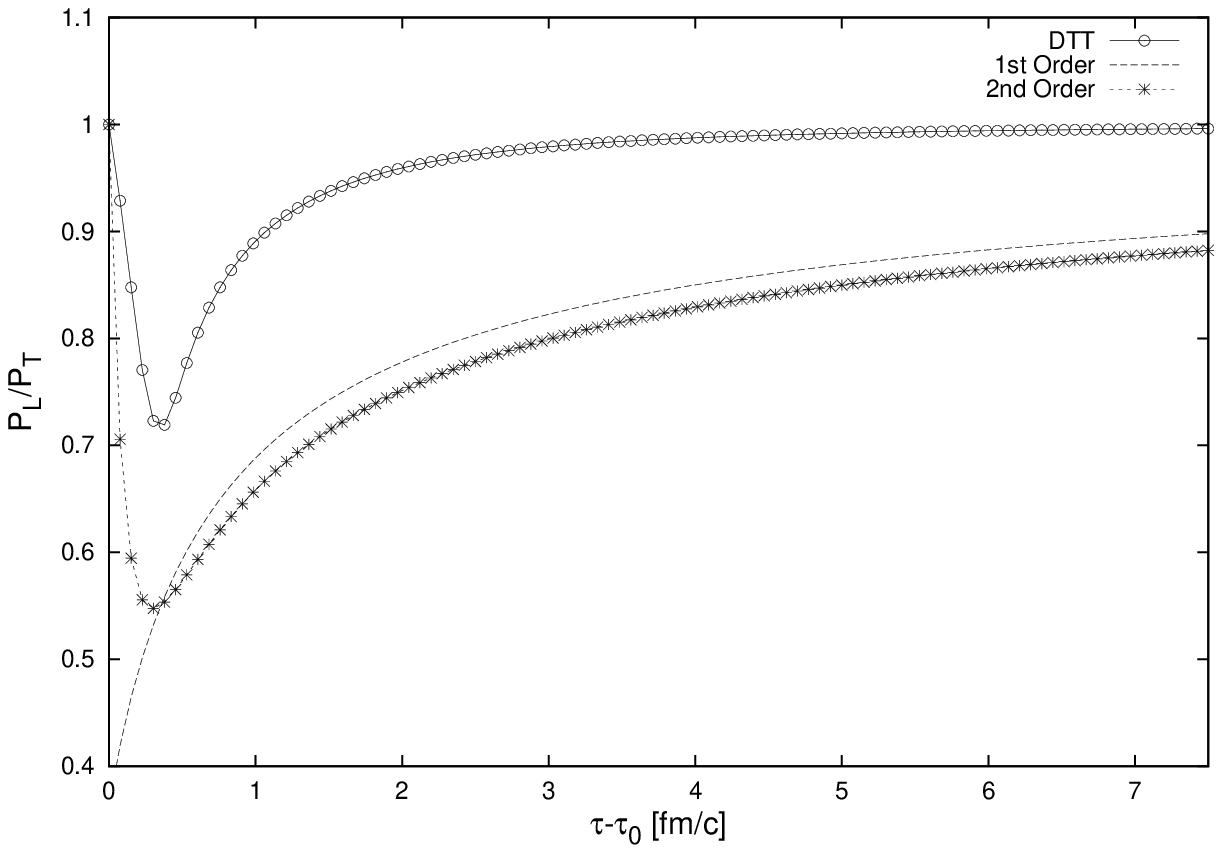}
\vspace{1cm}
\caption{Pressure isotropy $P_L/P_T$ as a function of proper time, for the DTT, second-order and Navier-Stokes theories with $\eta/s=0.09$.}
\label{compP1}
\end{figure}

\begin{figure}
\includegraphics{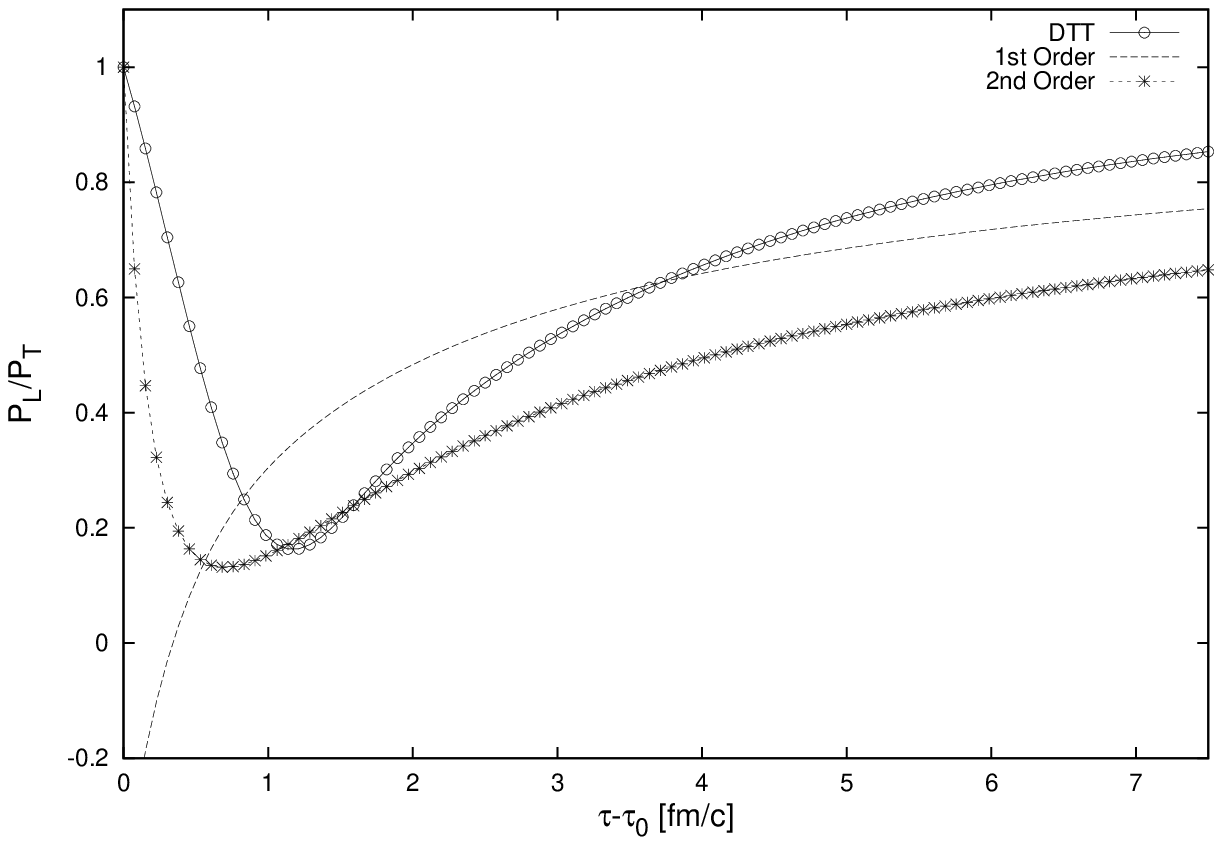}
\vspace{1cm}
\caption{Pressure isotropy $P_L/P_T$ as a function of proper time, for the DTT, second-order and Navier-Stokes theories with $\eta/s=0.375$.}
\label{compP2}
\end{figure}

In Figure \ref{compP1} we show the evolution of the pressure isotropy for $\eta/s=0.09$. 
It is clearly seen that the approach to ideal hydrodynamics is faster in the DTT, which also occurs with $\eta/s=0.375$ (Figure \ref{compP2}). 

We note that, with respect to the second-order theory, our results are in good agreement with those of previous studies \cite{muronga04,baier06,houv,ael,raj}. Taking into account the behaviour of the two quantities that we analyzed, we arrive at the important conclusion that the relaxation towards ideal hydrodynamics is faster in the DTT than in the second-order theory. This means that, as expected on theoretical grounds, the hydrodynamic evolution in the DTT is closer to that obtained from transport theory (see in particular the detailed comparison between Navier-Stokes, Israel-Stewart and covariant transport theory carried out by Houvinen and Molnar in Ref. \cite{houv}).

\section{Summary and conclusions}
\label{concsec}

In this work, we have studied the (nonlinear) hydrodynamical description of a conformal field within the theoretical framework of divergence-type theories. We proved that the theory we develop is causal (in a set of fluid states near equilibrium) and satisfies the Second Law exactly. Since it does not rely on gradient expansions, it goes beyond second-order (in velocity gradients) theories, thus being a closed theory. However, it is limited to the case where the second-order transport coefficients $\lambda_2$ and $\lambda_3$ vanish. For this case, we showed that the second-order stress-energy tensor constructed from conformal invariants \cite{sonhydro,logan,bat} can be consistently derived via an adiabatic expansion from the DTT.

As the most simple example, we have also obtained the hydrodynamic equations of the DTT for Bjorken flow, and compared them, analytically and numerically, with those of second-order and Navier-Stokes theories. The numerical calculations indicate that the relaxation towards ideal hydrodyanamics is substancially faster in the DTT as compared to the second-order theory. This indicates that the DTT is a better approximation to transport theory than the second-order theory, as expected since the former includes all-order velocity gradients. 

As stated in the Introduction, we think that the theory we have presented may be useful in the analysis of early-time dynamics and in the evolution of initial state fluctuations in heavy-ion collisions, essentially because the theory is not based on an expansion in velocity gradients. The extension of the DTT to include the case $\lambda_{2,3}\neq 0$ is also interesting. Work is in progress along these lines. 

\begin{acknowledgments}
We thank Paul Romatschke, Robert Geroch and Dirk Rischke for valuable comments and suggestions. This work has been supported in part by ANPCyT, CONICET and UBA (Argentina).
\end{acknowledgments}

\end{document}